\documentclass[12pt,a4paper]{elsarticle}
\usepackage{ifthen} 
\usepackage{bm}
\usepackage{amssymb}
\usepackage{amsfonts}
\usepackage{upgreek}
\usepackage{xspace}
\usepackage{color}
\usepackage{amsmath}

\usepackage{lineno}  

\usepackage{graphicx}  
\usepackage{subfigure}

\newboolean{pdflatex}
\setboolean{pdflatex}{true} 
\newboolean{articletitles}
\setboolean{articletitles}{true}
\newboolean{uprightparticles}
\setboolean{uprightparticles}{false} 

\usepackage{hyperref}
\usepackage[all]{hypcap}

\def\lhcb {LHCb\xspace}
\def\Pmu         {\ensuremath{\mu}\xspace}                 
\def\Ppsi        {\ensuremath{\psi}\xspace}                 
\def\PJ      {\ensuremath{J}\xspace}                 
\def\Pc      {\ensuremath{c}\xspace}                 
\def\Pchi        {\ensuremath{\chi}\xspace}                 
\def\Ppsi        {\ensuremath{\psi}\xspace}                 
\def\jpsi     {\ensuremath{{\PJ\mskip -3mu/\mskip -2mu\Ppsi\mskip 2mu}}\xspace}
\def\psitwos  {\ensuremath{\Ppsi{(2S)}}\xspace}
\def\chiczero {\ensuremath{\Pchi_{\cquark 0}}\xspace}
\def\chicone  {\ensuremath{\Pchi_{\cquark 1}}\xspace}
\def\chictwo  {\ensuremath{\Pchi_{\cquark 2}}\xspace}
\def\chic  {\ensuremath{\Pchi_{c}}\xspace}
\def\cquark    {\ensuremath{\Pc}\xspace}
\def\cquarkbar {\ensuremath{\overline \cquark}\xspace}
\def\ccbar     {\ensuremath{\cquark\cquarkbar}\xspace}
\def\mup        {\ensuremath{\Pmu^+}\xspace}
\def\mun        {\ensuremath{\Pmu^-}\xspace} 
\def\BF         {{\ensuremath{\cal B}\xspace}}
\def\BR         {\BF}
\newcommand{\tev}{\ensuremath{\mathrm{\,Te\kern -0.1em V}}\xspace}
\newcommand{\mevc}{\ensuremath{{\mathrm{\,Me\kern -0.1em V\!/}c}}\xspace}
\newcommand{\gevc}{\ensuremath{{\mathrm{\,Ge\kern -0.1em V\!/}c}}\xspace}
\newcommand{\mevcc}{\ensuremath{{\mathrm{\,Me\kern -0.1em V\!/}c^2}}\xspace}
\newcommand{\gevcc}{\ensuremath{{\mathrm{\,Ge\kern -0.1em V\!/}c^2}}\xspace}
\def\invpb {\ensuremath{\mbox{\,pb}^{-1}}\xspace}
\def\ps   {\ensuremath{{\rm \,ps}}\xspace}
\def\evtgen     {\mbox{\textsc{EvtGen}}\xspace}
\def\pythia     {\mbox{\textsc{Pythia}}\xspace}
\def\geant      {\mbox{\textsc{Geant4}}\xspace}

\def\photos     {\mbox{\textsc{Photos}}\xspace}
\def\chigen     {\mbox{\textsc{ChiGen}}\xspace}
\newcommand{\ie}{\mbox{\itshape i.e.}}

\newcommand{\Reference}{Ref.}
\newcommand{\References}{Refs.}
\newcommand{\Figure}{Fig.}
\newcommand{\Table}{Table}
\newcommand{\Section}{Sect.}
\newcommand{\Equation}{Eq.}
\newcommand{\MeVc}{\ensuremath{\mathrm{MeV}/c}}
\newcommand{\MeVcc}{\ensuremath{\mathrm{MeV}/c^{2}}}
\newcommand{\GeVc}{\ensuremath{\mathrm{GeV}/c}}
\newcommand{\GeVcc}{\ensuremath{\mathrm{GeV}/c^{2}}}
\newcommand{\TeV}{\ensuremath{\mathrm{TeV}}}

\newcommand{\pp}{\ensuremath{pp}}
\newcommand{\ppbar}{\ensuremath{p\bar{p}}}
\newcommand{\Jpsi}{\jpsi}
\newcommand{\Chic}{\Pchi\xspace} 
\newcommand{\PsiTwoS}{\psitwos}
\newcommand{\ChicJ}{\ensuremath{\chi_{cJ}}}
\newcommand{\ChicJOneP}{\ensuremath{\ChicJ(1P)}}
\newcommand{\ChicZero}{\chi_{c0}}
\newcommand{\ChicOne}{\chi_{c1}}
\newcommand{\ChicTwo}{\chi_{c2}}
\newcommand{\tz}{\ensuremath{t_{z}}}
\newcommand{\SqrtS}{\ensuremath{\sqrt{s}\myop{=}\myvalue{7}{\TeV}}}
\newcommand{\LHCb}{LHCb}
\newcommand{\pT}{\ensuremath{p_{\mathrm{T}}}}
\newcommand{\pTJpsi}{\ensuremath{\pT^{\Jpsi}}}
\newcommand{\pTGamma}{\ensuremath{\pT^{\gamma}}}
\newcommand{\pGamma}{\ensuremath{p^{\gamma}}}
\newcommand{\CLgamma}{\ensuremath{\mathrm{CL}_{\gamma}}}
\newcommand{\DeltaM}{\ensuremath{\Delta M}}
\newcommand{\pTJpsiRange}{\ensuremath{2\myop{<}\pTJpsi\myop{<}15~\GeVc}}
\newcommand{\RapidityJpsi}{\ensuremath{y^{\Jpsi}}}
\newcommand{\yRange}{\ensuremath{2.0\myop{<}\RapidityJpsi\myop{<}4.5}}
\newcommand{\etaRange}{\ensuremath{2 \myop{<}\eta \myop{<}5}}
\newcommand{\MuMu}{\ensuremath{\mup \mun}}
\newcommand{\JpsiToMuMu}{\ensuremath{\Jpsi \rightarrow \mup \mun}}
\newcommand{\PsiTwoSToJpsiX}{\ensuremath{\PsiTwoS \rightarrow \Jpsi X}}
\newcommand{\ChicToJpsiGamma}{\ensuremath{\Chic \rightarrow \Jpsi \gamma}}
\newcommand{\ChicJToJpsiGamma}{\ensuremath{\ChicJ \rightarrow \Jpsi \gamma}}
\newcommand{\ChicZeroToJpsiGamma}{\ensuremath{\ChicZero \rightarrow \Jpsi \gamma}}
\newcommand{\ChicOneToJpsiGamma}{\ensuremath{\ChicOne \rightarrow \Jpsi \gamma}}
\newcommand{\BToJpsiK}{\ensuremath{B^+\rightarrow \Jpsi K^+}}
\newcommand{\BToChicK}{\ensuremath{B^+\rightarrow \Chic K^+}}
\newcommand{\BToChicOneK}{\ensuremath{B^+\rightarrow \ChicOne K^+}}
\newcommand{\brBToJpsiK}{\BR(\BToJpsiK)}
\newcommand{\brBToChicK}{\BR(\BToChicK)}
\newcommand{\brBToChicOneK}{\BR(\BToChicOneK)}
\newcommand{\brBToChicOneKmeas}{(4.6\pm 0.4)\times 10^{-4}}
\newcommand{\brChicOneToJpsiGamma}{\BR(\ChicOneToJpsiGamma)}
\newcommand{\brChicOneToJpsiGammameas}{(34.4\pm 1.5)\times 10^{-2}}
\newcommand{\brBToJpsiKmeas}{(1.013\pm 0.034)\times 10^{-3}}
\newcommand{\brChicToJpsiGamma}{\BR(\ChicToJpsiGamma)}
\newcommand{\brChicJToJpsiGamma}{\BR(\ChicJToJpsiGamma)}
\newcommand{\yBToJpsiKmeas}{\ensuremath{8440\pm 96}}
\newcommand{\yBToChicKmeas}{\ensuremath{142\pm 15}}
\newcommand{\srChicToJpsi}{\ensuremath{\sigma(\ChicToJpsiGamma)\,/\,\sigma(\Jpsi)}}
\newcommand{\myop}[1]{\ensuremath{\,{#1}\,}}
\newcommand{\myrange}[2]{\ensuremath{{#1}\myop{-}{#2}}}
\newcommand{\myvalue}[2]{\mbox{\ensuremath{{#1}\:{#2}}}} 

\renewcommand{\MeVc}{\mevc}
\renewcommand{\MeVcc}{\mevcc}
\renewcommand{\GeVc}{\gevc}
\renewcommand{\GeVcc}{\gevcc}
\renewcommand{\TeV}{\tev}

\renewcommand{\Jpsi}{\jpsi}
\renewcommand{\Chic}{\chic}
\renewcommand{\ChicZero}{\chiczero}
\renewcommand{\ChicOne}{\chicone}
\renewcommand{\ChicTwo}{\chictwo}
\renewcommand{\LHCb}{\lhcb}

\usepackage{multirow}
\usepackage{afterpage}
\usepackage{mciteplus}

\begin{document}


\begin{titlepage}
\pagenumbering{roman}

\vspace*{-1.5cm}
\centerline{\large EUROPEAN ORGANIZATION FOR NUCLEAR RESEARCH (CERN)}
\vspace*{1.5cm}
\hspace*{-0.5cm}
\begin{tabular*}{\linewidth}{lc@{\extracolsep{\fill}}r}
\ifthenelse{\boolean{pdflatex}}
{\vspace*{-2.7cm}\mbox{\!\!\!\includegraphics[width=.14\textwidth]{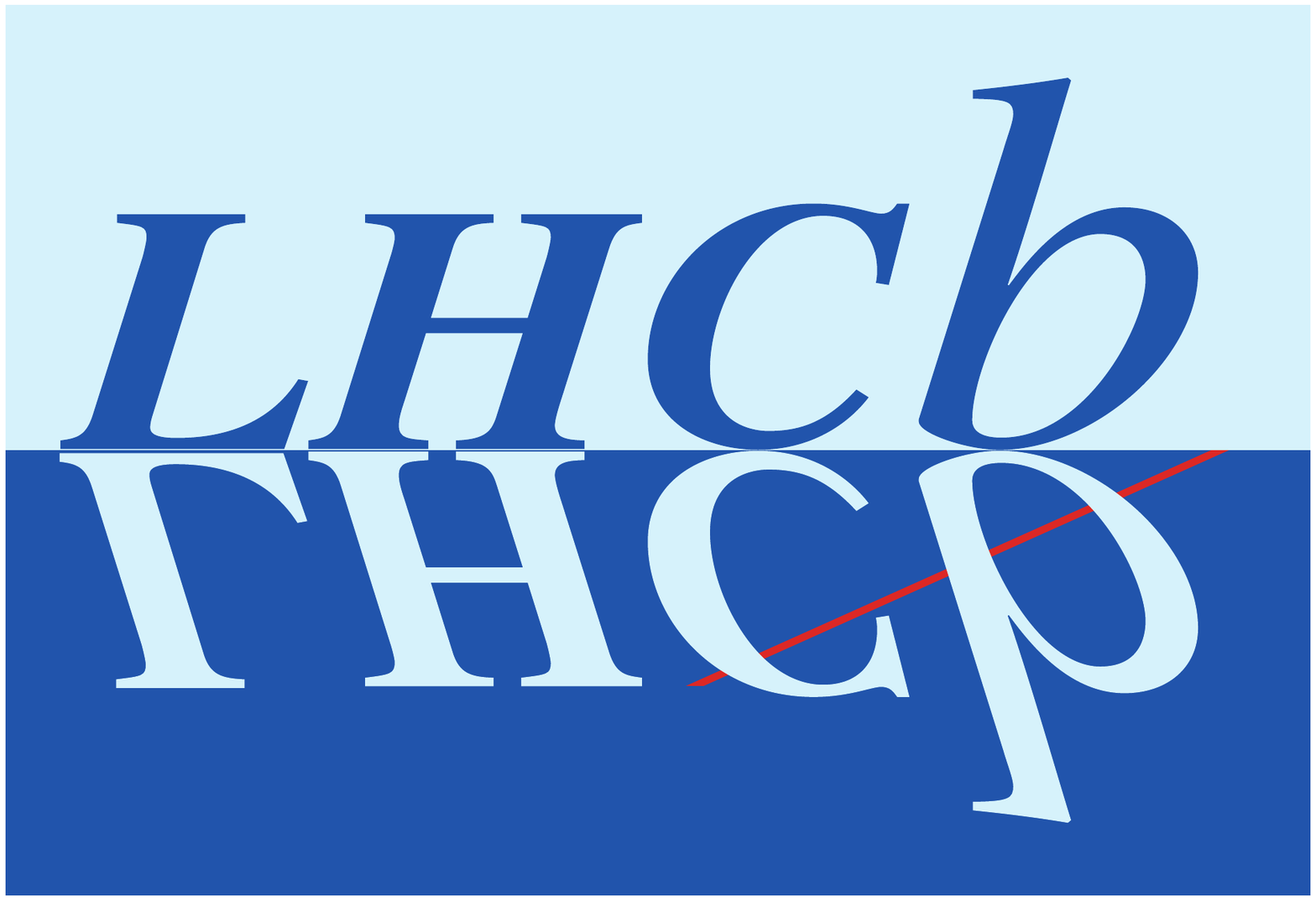}} & &}%
{\vspace*{-1.2cm}\mbox{\!\!\!\includegraphics[width=.12\textwidth]{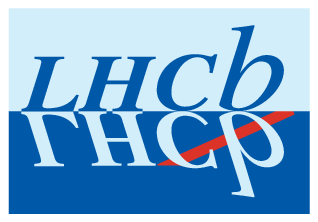}} & &}%
\\
 & & CERN-PH-EP-2012-068 \\  
 & & LHCb-PAPER-2011-030 \\  
 & & \today \\ 
 & & \\
\end{tabular*}

\vspace*{4.0cm}

{\bf\boldmath\huge
\begin{center}
Measurement of the ratio of prompt \Chic\ to \Jpsi\ production 
in \pp\ collisions at \SqrtS\ 
\end{center}
}

\vspace*{2.0cm}

\begin{center}
The LHCb collaboration\footnote{Authors are listed on the following pages.}
\end{center}

\vspace{\fill}

\begin{abstract}
  \noindent 
  The prompt production of charmonium 
  $\chi_{c}$ and $J/\psi$ states is studied in
  proton-proton collisions at a
  centre-of-mass energy of $\sqrt{s}=7$ TeV at the Large Hadron Collider. 
  The $\chi_{c}$ and $J/\psi$ mesons are identified through
  their decays $\chi_{c}\rightarrow J/\psi \gamma$ and 
  $J/\psi\rightarrow \mu^+\mu^-$ using
  36 pb$^{-1}$ of data collected by the LHCb detector in 2010. 
  The ratio of the prompt production cross-sections for $\chi_{c}$ 
  and $J/\psi$,
  $\sigma (\chi_{c}\rightarrow J/\psi \gamma )/ \sigma (J/\psi)$,
  is determined as a function of the $J/\psi$ 
  transverse momentum in the range 
  $2 < p_{\mathrm T}^{J/\psi} < 15$ GeV/$c$. 
  The results are in excellent agreement with next-to-leading order
  non-relativistic expectations and 
  show a significant discrepancy compared with the 
  colour singlet model prediction at leading order, 
  especially in the low $p_{\mathrm T}^{J/\psi}$ region.
\end{abstract}

\vspace*{1.0cm}

\begin{center}
\it{Submitted to Phys. Lett. B}
\end{center}

\vspace{\fill}

\end{titlepage}

\maketitle

\newpage
\setcounter{page}{2}
\mbox{~}
\newpage

\centerline{\large\bf LHCb collaboration}
\begin{flushleft}
\small
R.~Aaij$^{38}$, 
C.~Abellan~Beteta$^{33,n}$, 
B.~Adeva$^{34}$, 
M.~Adinolfi$^{43}$, 
C.~Adrover$^{6}$, 
A.~Affolder$^{49}$, 
Z.~Ajaltouni$^{5}$, 
J.~Albrecht$^{35}$, 
F.~Alessio$^{35}$, 
M.~Alexander$^{48}$
% ...

\footnotesize{\it
% ...
$ ^{5}$Clermont Universit\'{e}, Universit\'{e} Blaise Pascal, CNRS/IN2P3, LPC, Clermont-Ferrand, France\\
$ ^{6}$CPPM, Aix-Marseille Universit\'{e}, CNRS/IN2P3, Marseille, France\\
% ...
$ ^{33}$Universitat de Barcelona, Barcelona, Spain\\
$ ^{34}$Universidad de Santiago de Compostela, Santiago de Compostela, Spain\\
$ ^{35}$European Organization for Nuclear Research (CERN), Geneva, Switzerland\\
% ...
$ ^{38}$Nikhef National Institute for Subatomic Physics, Amsterdam, The Netherlands\\
% ...
$ ^{43}$H.H. Wills Physics Laboratory, University of Bristol, Bristol, United Kingdom\\
% ...
$ ^{48}$School of Physics and Astronomy, University of Glasgow, Glasgow, United Kingdom\\
$ ^{49}$Oliver Lodge Laboratory, University of Liverpool, Liverpool, United Kingdom\\
% ...
$ ^{n}$LIFAELS, La Salle, Universitat Ramon Llull, Barcelona, Spain\\
% ...
}
\clearpage

\end{flushleft}

\cleardoublepage


\pagestyle{plain} 
\setcounter{page}{1}
\pagenumbering{arabic}

\section{Introduction}
\label{sec:Introduction}

The study of charmonium production provides an important 
test of the underlying mechanisms
described by Quantum Chromodynamics (QCD).
At the centre-of-mass energies of
proton-proton collisions at the Large Hadron Collider, 
\ccbar\ pairs are expected
to be produced predominantly via Leading Order (LO) gluon-gluon interactions, 
followed by the formation of bound charmonium states. 
The former can be calculated using perturbative QCD and
the latter is described by non-perturbative models. 
Other, more recent, approaches make
use of non-relativistic QCD factorization (NRQCD),
which assumes the \ccbar pair to be a combination of 
colour-singlet and colour-octet states 
as it evolves towards the final bound
system via the exchange of soft gluons \cite{Bodwin:1994jh}.
The fraction of \Jpsi\ produced through the radiative decay of
\Chic\ states is an important test of both the
colour-singlet and colour-octet production mechanisms.
In addition, knowledge of this fraction is required 
for the measurement of the 
\Jpsi\ polarisation,
since the predicted polarisation is different for \Jpsi\ mesons
coming from the radiative decay of \Chic\ state compared to those
that are directly produced. 

In this paper, we report the measurement 
of the ratio of the cross-sections for the production of 
$P$-wave charmonia \ensuremath{\ChicJOneP}, with $J=$ 0, 1, 2, 
to the production of \Jpsi\ in promptly produced charmonium.
The ratio is measured 
as a function of the \Jpsi\ transverse momentum 
in the range \pTJpsiRange\ and in the
rapidity range \yRange.
Throughout the paper we refer to the collection 
of \ensuremath{\ChicJOneP} states as \Chic. 
The \Chic\ and \Jpsi\ candidates 
are reconstructed through their respective decays
\ChicToJpsiGamma\ and \JpsiToMuMu\
using a data sample corresponding to an integrated luminosity of 36~\invpb\
collected during 2010.
Prompt (non-prompt)
production refers to charmonium states produced at the
interaction point (in the decay of $b$-hadrons); 
direct production refers to prompt \Jpsi\ mesons 
that are not decay products of 
an intermediate resonant state, such as the \PsiTwoS. 
The measurements are complementary to the 
measurements of the \Jpsi\ production 
cross-section~\cite{Aaij:2011jh}
and the ratio of the prompt \Chic\ production cross-sections
for the $J=1$ and $J=2$ spin states~\cite{Aaij:2011chic},  
and extend the \pTJpsi\ coverage with 
respect to previous experiments~\cite{Abt:2008ed,Abe:1997yz}.

\section{\LHCb\ detector and selection requirements}
\label{sec:Detector}

The \LHCb\ detector~\cite{Alves:2008zz} is a single-arm forward spectrometer
with a pseudo-rapidity range \etaRange.
The detector consists of a silicon vertex detector, 
a dipole magnet, 
a tracking system, 
two ring-imaging Cherenkov (RICH) detectors, a calorimeter
system and a muon system.

Of particular importance in this measurement are the calorimeter and muon
systems. 
The calorimeter system consists of a scintillating pad
detector (SPD) and a pre-shower system, followed by
electromagnetic (ECAL) and hadron calorimeters. 
The SPD and pre-shower 
are designed to distinguish between signals from photons and
electrons. 
The ECAL is constructed from scintillating tiles interleaved with lead tiles. 
Muons are identified using hits in muon chambers 
interleaved with iron filters. 

The signal simulation sample used for this analysis was generated using the
\pythia~\ensuremath{6.4} generator~\cite{Sjostrand:2006za} 
configured with the parameters detailed in \Reference~\cite{5873949}.
The \evtgen~\cite{Lange:2001uf}, \photos~\cite{Barberio:1993qi} 
and \geant~\cite{Agostinelli:2002hh} 
packages were used to decay unstable particles, 
generate QED radiative corrections and 
simulate interactions in the detector, respectively.
The sample consists of events in which at least one \JpsiToMuMu\ decay
takes place with no constraint on the production mechanism.

The trigger consists of a hardware stage followed by a software stage,
which applies a full event reconstruction. 
For this analysis, events are selected which have been
triggered by a
pair of oppositely charged muon candidates, where either one of the muons has a
transverse momentum \ensuremath{\pT\myop{>}\myvalue{1.8}{\GeVc}} or one of the
pair has \ensuremath{\pT\myop{>}\myvalue{0.56}{\GeVc}} and the other has
\ensuremath{\pT\myop{>}\myvalue{0.48}{\GeVc}}. 
The invariant mass of the
candidates is required to be greater than \myvalue{2.9}{\GeVcc}. The photons are
not involved in the trigger decision for this analysis.

Photons are reconstructed using the electromagnetic
calorimeter and identified using a
likelihood-based estimator,
\CLgamma, 
constructed from variables that rely on calorimeter and tracking
information. 
For example, in order to reduce the electron background, candidate photon
clusters are required not to be matched to the trajectory
of a track extrapolated from the tracking system to the
cluster position in the calorimeter.
For each photon candidate a value of \CLgamma, 
with a range between 0 (background-like) and 1 (signal-like), 
is calculated based on simulated signal and background samples.

The photons are classified as one of two types:  
those that have converted to electrons
in the material after the dipole magnet and those that have not. 
Converted photons are identified as clusters in the ECAL
with correlated activity in the SPD.
In order to account for the different energy resolutions of the two types
of photons,
the analysis is performed separately for 
converted and non-converted photons 
and the results are combined.
Photons that convert before the magnet 
require a different analysis strategy and are not considered here.
The photons used to reconstruct the \Chic\ candidates are
required to have a transverse momentum
\ensuremath{\pTGamma\myop{>}\myvalue{650}{\MeVc}}, 
a momentum \ensuremath{\pGamma\myop{>}\myvalue{5}{\GeVc}} and
\ensuremath{\CLgamma\myop{>}0.5};
the efficiency of the \CLgamma\ cut for photons from \Chic\ decays 
is \ensuremath{72\%}.

All \Jpsi\ candidates
are reconstructed using the decay \JpsiToMuMu.
The muon and \Jpsi\ identification criteria are identical to
those used in \Reference~\cite{Aaij:2011jh}: 
each track must be identified as a muon with
\ensuremath{\pT\myop{>}700\MeVc} and have a track fit
\ensuremath{\chi^{2}/\mathrm{ndf}\myop{<}4}, where \ensuremath{\mathrm{ndf}} is
the number of degrees of freedom. The two muons must originate from a
vertex with a probability of the vertex fit greater than
\ensuremath{0.005}.
In addition, the \MuMu\
invariant mass is required to be in the range
\myrange{3062}{\myvalue{3120}{\MeVcc}}.
The \Chic\ candidates are formed from the 
selected \Jpsi\ candidates and photons. 
 
The non-prompt \Jpsi\ contribution arising from $b$-hadron decays is
taken from \Reference~\cite{Aaij:2011jh}. 
For the \Chic\ candidates,
the \Jpsi\ pseudo-decay time, \tz, 
is used to reduce the contribution from non-prompt decays, 
by requiring
\ensuremath{\tz\myop{=}
(z_{\Jpsi}\myop{-}z_{PV})M_{\Jpsi}\myop{/}p_{z}<\myvalue{0.1}{\ps}},
where \ensuremath{M_{\Jpsi}} is the reconstructed dimuon invariant mass,
\ensuremath{z_{\Jpsi}-z_{\mathrm{PV}}} is the
\ensuremath{z} separation of the reconstructed
production (primary) and decay vertices of the dimuon,
and \ensuremath{p_{z}} is the \ensuremath{z}-component 
of the dimuon momentum. 
The \ensuremath{z}-axis is
parallel to the beam line in the centre-of-mass frame. 
Simulation studies show that, with this requirement
applied, the remaining fraction of \Chic\ from $b$-hadron decays is about
\ensuremath{0.1\%}. 
This introduces an uncertainty much smaller than any of the
other systematic or statistical uncertainties 
evaluated in this analysis and is not considered further. 

\begin{figure}[!htbp]
  \begin{center}
    \subfigure{
      \ifthenelse{\boolean{pdflatex}}{
	\includegraphics*[width=0.48\textwidth]{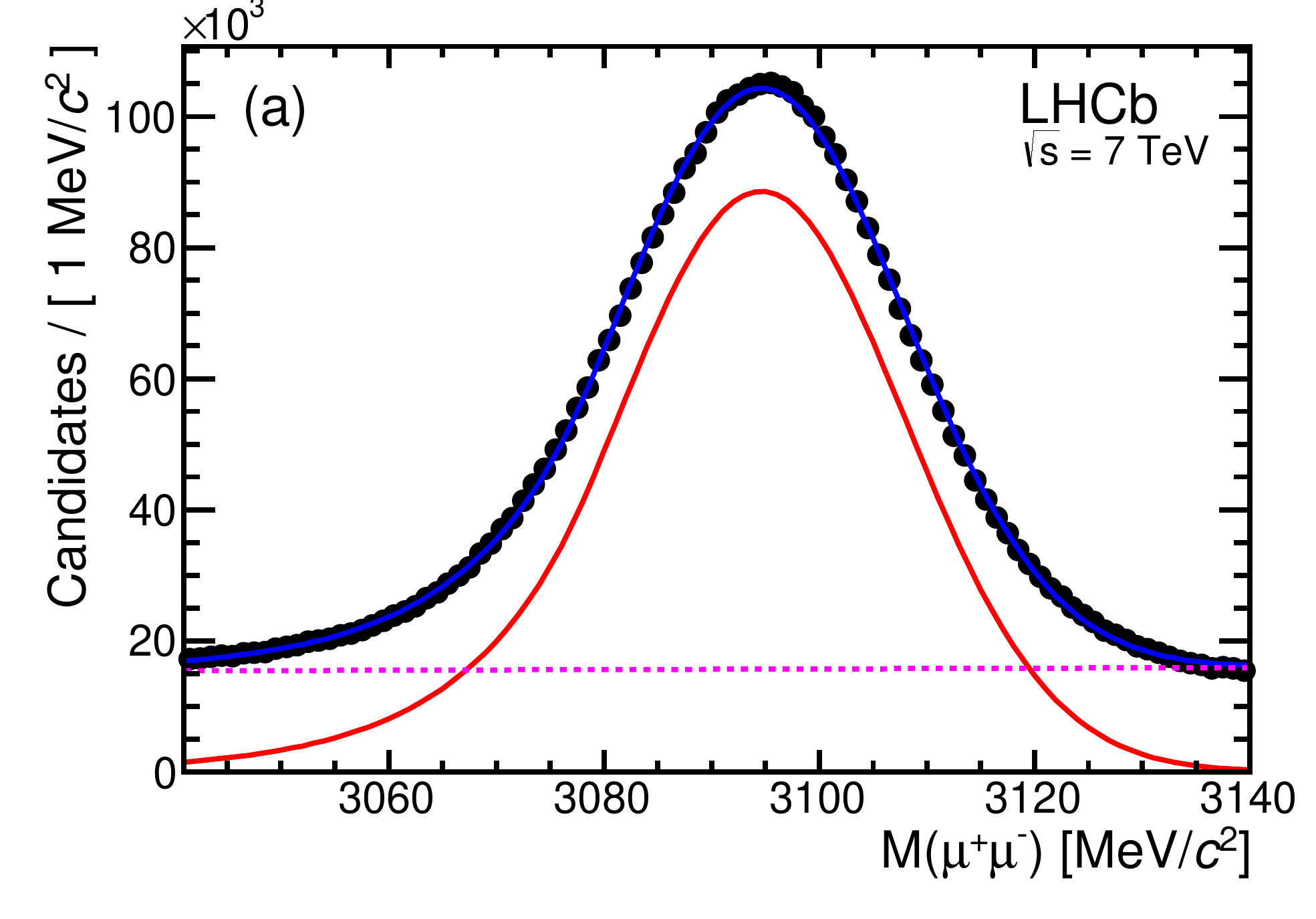}
      }{
	\includegraphics*[width=0.48\textwidth]{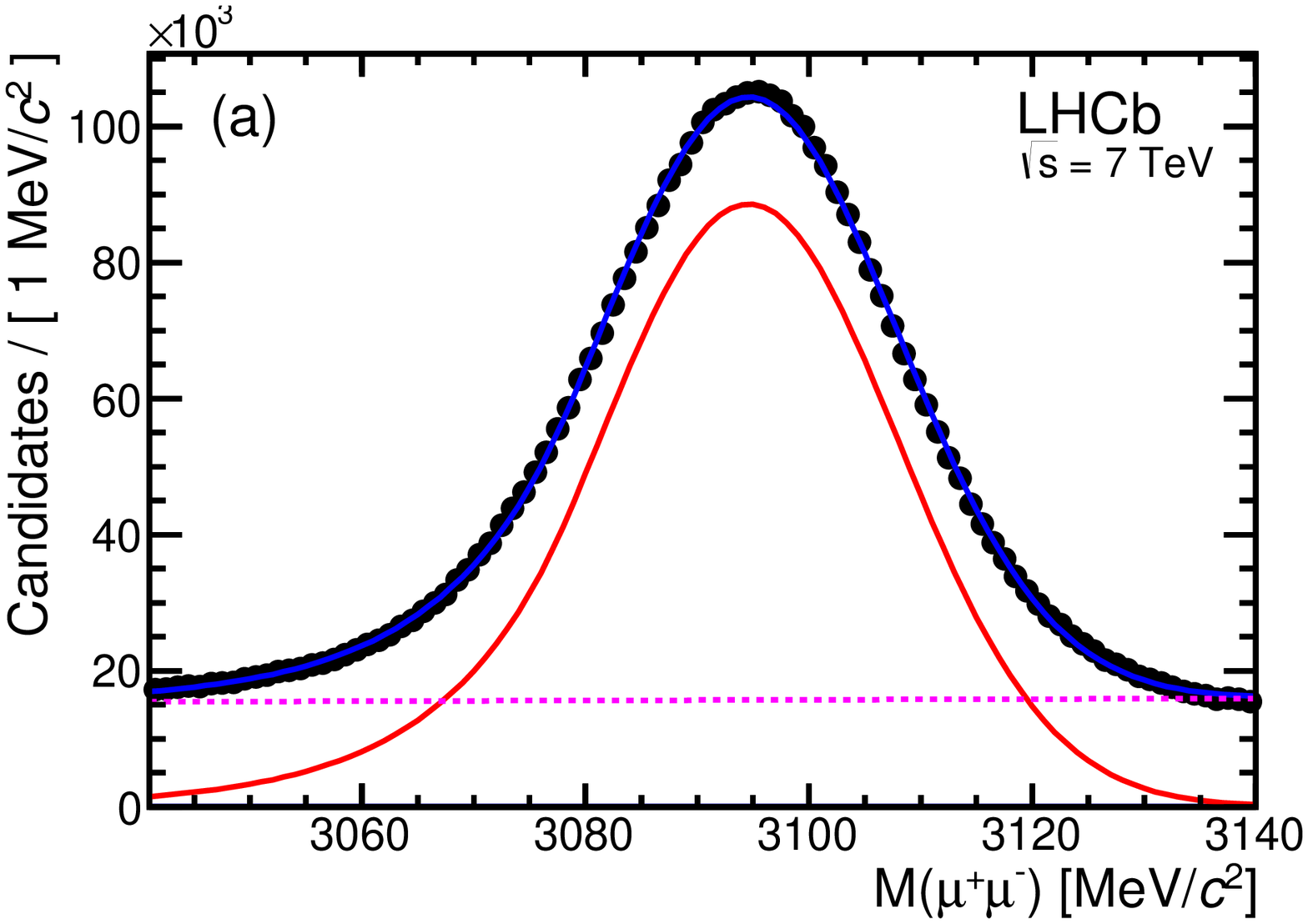}}}
    \subfigure{
      \ifthenelse{\boolean{pdflatex}}{
	\includegraphics*[width=0.48\textwidth]{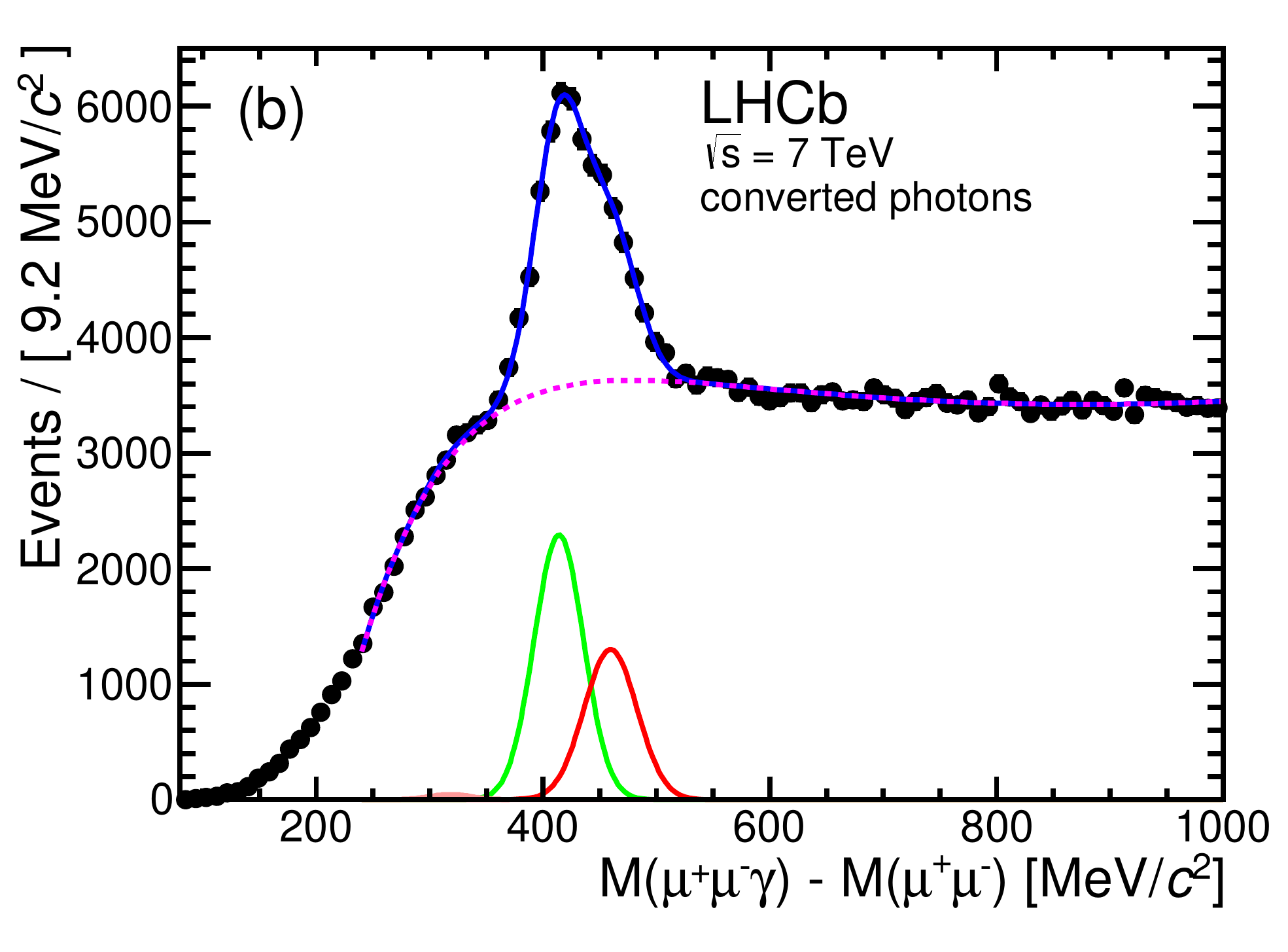}
      }{
	\includegraphics*[width=0.48\textwidth]{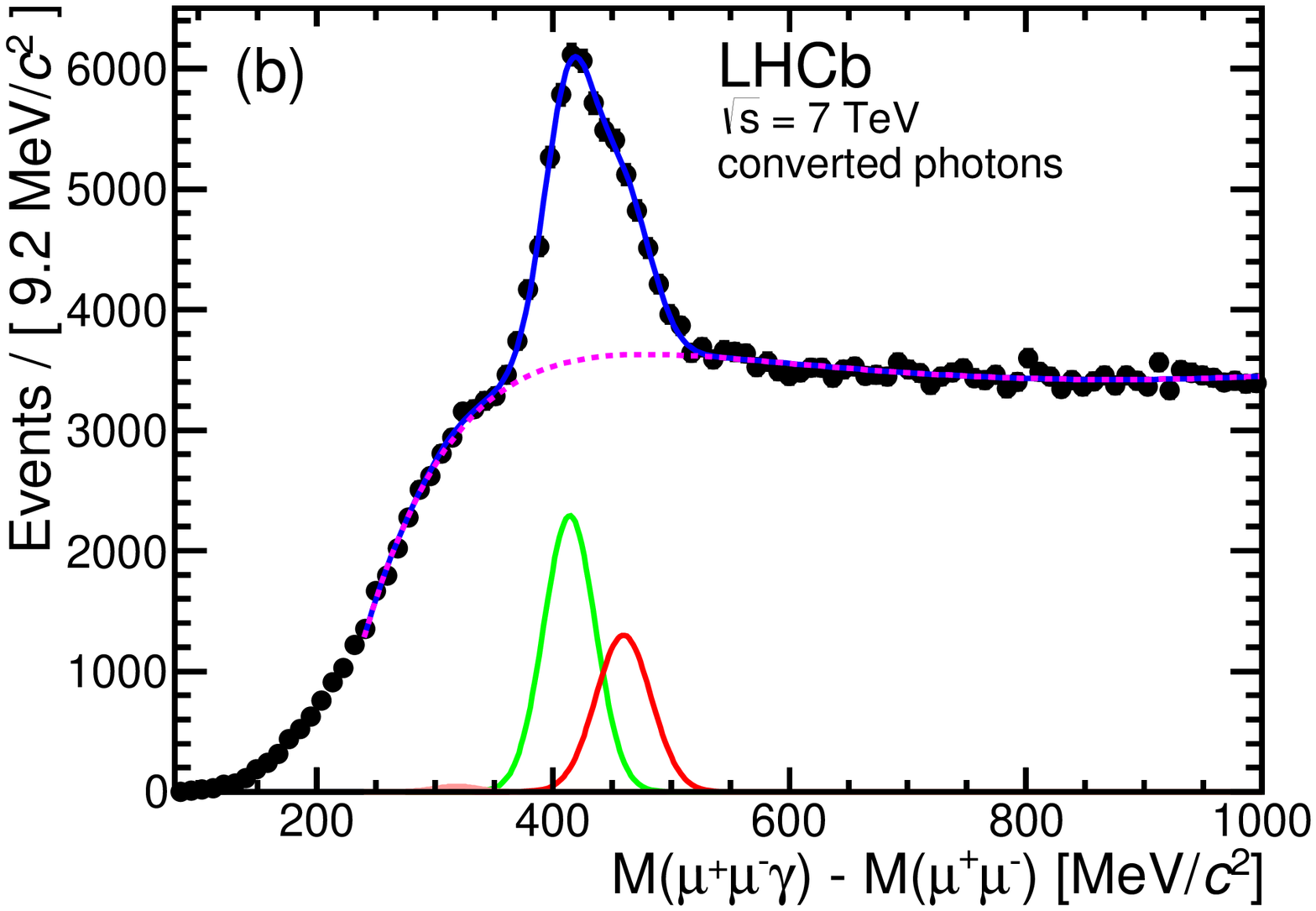}}}
    \subfigure{
      \ifthenelse{\boolean{pdflatex}}{
	\includegraphics*[width=0.48\textwidth]{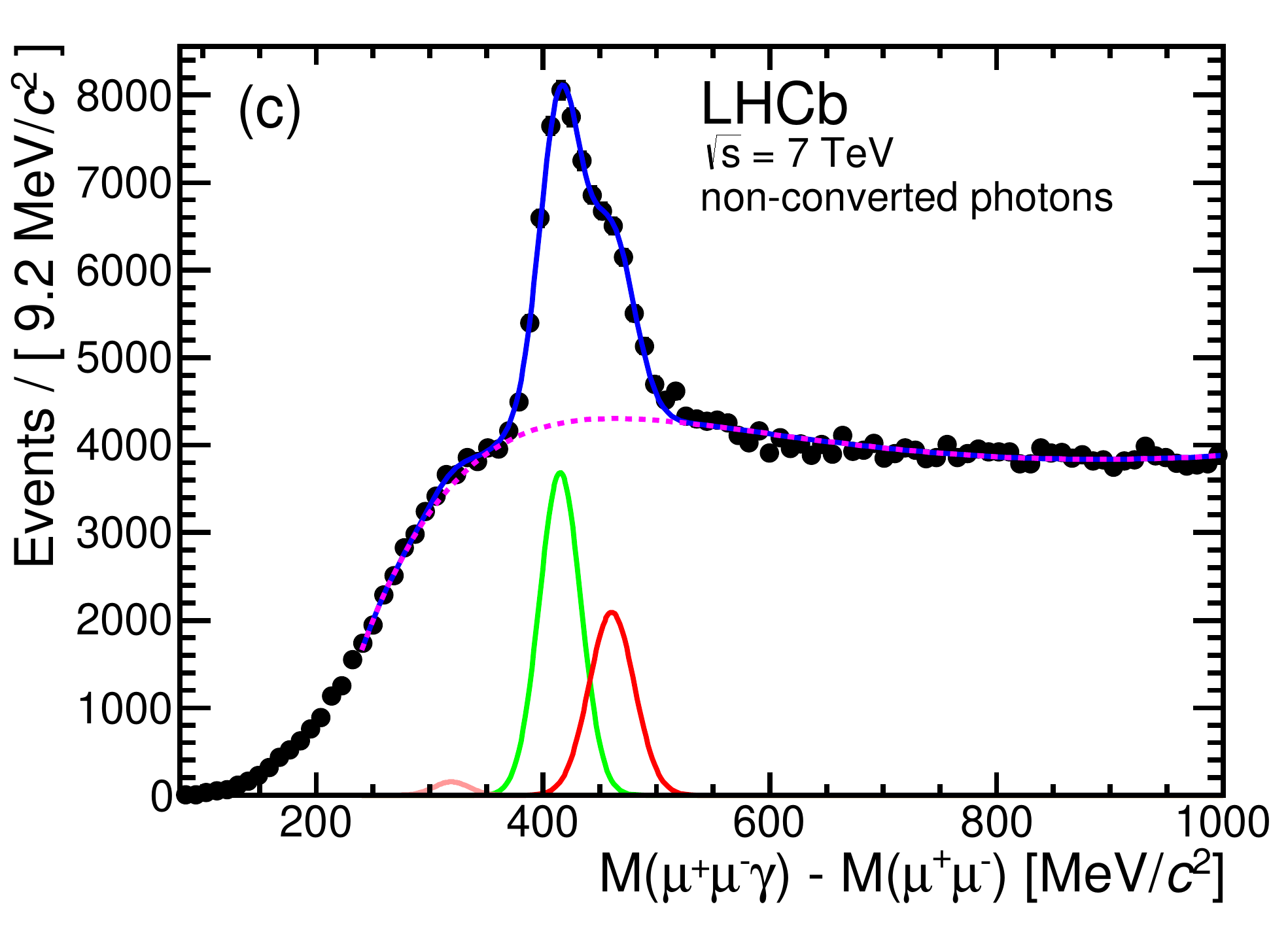}
      }{
	\includegraphics*[width=0.48\textwidth]{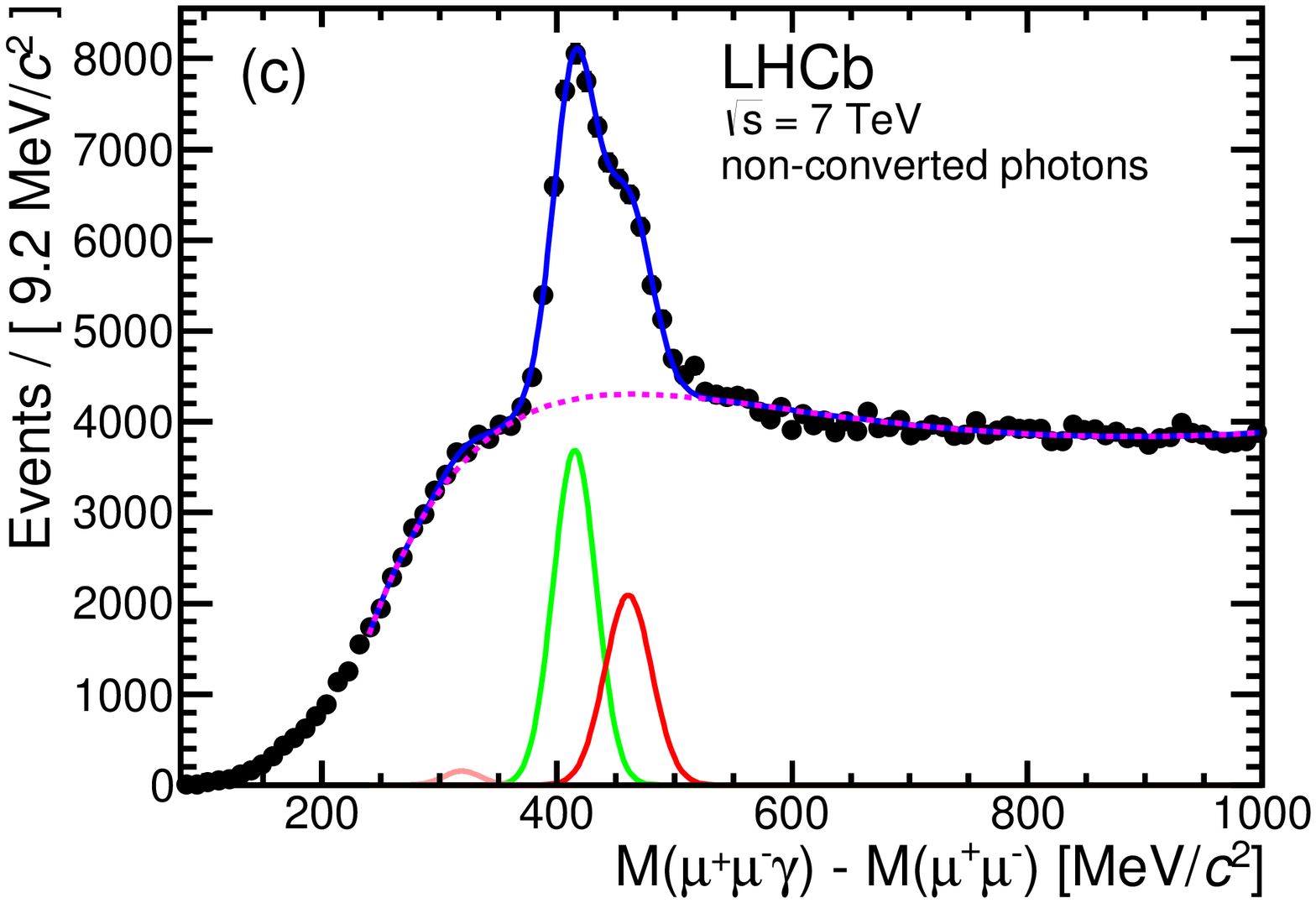}}}
    \caption{\small{(a) Invariant mass of the \MuMu\ pair for 
	selected \Jpsi\ candidates. 
	The solid red curve corresponds to the signal and
	the background is shown as a dashed purple curve.
	(b) and (c) show the 
	\ensuremath{\DeltaM\myop{=}M\left(\mup \mun\,\gamma\right)\myop{-}M\left(\mup \mun\right)}
	distributions of selected \Chic\ candidates 
	with (b) converted and (c) non-converted
	photons.
	The upper solid blue curve corresponds to the overall fit function
	described in \Reference~\cite{Aaij:2011chic}. 
	The lower solid curves correspond to the fitted \ChicZero,
	\ChicOne\ and \ChicTwo\ contributions from left to right, respectively
	(the \ChicZero\ peak is barely visible). 
	The background distribution is
	shown as a dashed purple curve.}}
    \label{fig:MassPlots}
  \end{center}
\end{figure}

The distributions of the 
\MuMu\ mass of selected \Jpsi\ candidates and
the mass difference,
\ensuremath{\DeltaM\myop{=}M\left(\mup \mun\,\gamma\right)\myop{-}M\left(\mup \mun\right)},
of the selected \Chic\ candidates for the converted and non-converted samples
are shown in \Figure~\ref{fig:MassPlots}. 
The total number of prompt \Jpsi\ candidates observed in the data
is \ensuremath{\sim 2.6} million.
The fit procedure to extract the three \Chic\ signal yields using Gaussian functions 
and one common function for the combinatorial background
is discussed in \Reference~\cite{Aaij:2011chic}.
The total number of \ChicZero, \ChicOne\ and \ChicTwo\ candidates observed
are \ensuremath{823}, \ensuremath{38\,630} 
and \ensuremath{26\,114} respectively.
Since the \ChicZeroToJpsiGamma\ branching fraction is
\ensuremath{\sim 30} (17) times 
smaller than that of the \ChicOne\ (\ChicTwo), 
the yield of \ChicZero\ is small as expected~\cite{Nakamura:2010zzi}.

\section{Determination of the cross-section ratio}
\label{sec:ExpMethod}

\afterpage{\clearpage}
The main contributions to the production of prompt \Jpsi\ arise from direct production 
and from the feed-down processes \ChicToJpsiGamma\ and \PsiTwoSToJpsiX\ where $X$ refers to
any final state.  
The cross-section ratio for the production of prompt \Jpsi\ from \ChicToJpsiGamma\ decays
compared to all prompt \Jpsi\ can be expressed in terms of the three
\ChicJ\ $(J=0,1,2)$ signal yields,
\ensuremath{N_{\ChicJ}},
and the prompt \Jpsi\ yield,
\ensuremath{N_{\Jpsi}}, as
\begin{align}
\dfrac{\sigma(\ChicToJpsiGamma)}{\sigma(\Jpsi)} 
& \, \approx \,  
\dfrac{\sigma(\ChicToJpsiGamma)}
{\sigma^{\mathrm{dir}}(\Jpsi)+\sigma(\PsiTwoSToJpsiX)+\sigma(\ChicToJpsiGamma)} \nonumber \\ 
&  \nonumber \\ 
& \, \myop{=} \, 
\dfrac
{
\displaystyle{\sum_{J=0}^{J=2}}\;
\dfrac{N_{\ChicJ}}{\epsilon^{\ChicJ}_{\gamma} \epsilon^{\ChicJ}_{\mathrm{sel}} }
\cdot
\dfrac{\epsilon^{\mathrm{dir}}_{\Jpsi}}{\epsilon^{\ChicJ}_{\Jpsi}}
}
{
N_{\Jpsi}R_{2S}
+
\displaystyle{\sum_{J=0}^{J=2}}\;
\dfrac{N_{\ChicJ}}{\epsilon^{\ChicJ}_{\gamma} \epsilon^{\ChicJ}_{\mathrm{sel}}}
\left[ 
\dfrac{\epsilon^{\mathrm{dir}}_{\Jpsi}}{\epsilon^{\ChicJ}_{\Jpsi}} 
-R_{2S}\right]
} \label{e:ChiC_Jpsi_ratio}\\
\intertext{with}
R_{2S} & \, \myop{=} \, 
\dfrac{1+f_{2S}}
{1+f_{2S}\dfrac{\epsilon_{\Jpsi}^{2S}}{\epsilon_{\Jpsi}^{\mathrm{dir}}}} \label{e:ChiC_Jpsi_ratioR} \\
\intertext{and}
f_{2S} & \, \myop{=} \, \dfrac{\sigma(\PsiTwoSToJpsiX)}{\sigma^{\mathrm{dir}}(\Jpsi)}.
\end{align}
The total prompt \ChicToJpsiGamma\ cross-section is 
\ensuremath{\sigma (\ChicToJpsiGamma)=
\sum_{J=0}^{J=2} \; \sigma_{\ChicJ}\cdot \brChicJToJpsiGamma}
where \ensuremath{\sigma_{\ChicJ}} is the production cross-section
for each \ChicJ\ state and \brChicJToJpsiGamma\ is the 
corresponding branching fraction. 
The cross-section ratio $f_{2S}$ 
is used to link the prompt \PsiTwoS\ contribution to the
direct \Jpsi\ contribution and
$R_{2S}$ takes into account their efficiencies.
The combination of the trigger,
reconstruction and selection efficiencies
for direct \Jpsi,
for \Jpsi\ from \PsiTwoS\ decay, and
for \Jpsi\ from \ChicToJpsiGamma\ decay are
\ensuremath{\epsilon^{\mathrm{dir}}_{\Jpsi}},  
\ensuremath{\epsilon^{2S}_{\Jpsi}}, and
\ensuremath{\epsilon^{\ChicJ}_{\Jpsi}} respectively.
The efficiency to reconstruct and select a photon
from a \ChicToJpsiGamma\ decay,
once the \Jpsi\ is already selected,
is \ensuremath{\epsilon^{\ChicJ}_{\gamma}} 
and the efficiency for the subsequent selection of the \ChicJ\ is
\ensuremath{\epsilon^{\ChicJ}_{\mathrm{sel}}}.

The efficiency terms in \Equation~(1) are determined using simulated events
and are partly validated with control channels in the data.
The results for the efficiency ratios 
\ensuremath{\epsilon^{2S}_{\Jpsi}/\epsilon^{\mathrm{dir}}_{\Jpsi}},
\ensuremath{\epsilon^{\mathrm{dir}}_{\Jpsi}/\epsilon^{\ChicJ}_{\Jpsi}}
and the product 
\ensuremath{\epsilon^{\ChicJ}_{\gamma}\epsilon^{\ChicJ}_{\mathrm{sel}}}
are discussed in \Section~\ref{sec:Efficiencies}.

The prompt \ensuremath{N_{\Jpsi}} and \ensuremath{N_{\ChicJ}} yields
are determined in bins of \pTJpsi\ in the range \pTJpsiRange\
using the methods described in 
\References~\cite{Aaij:2011jh} and~\cite{Aaij:2011chic} respectively.
In \Reference~\cite{Aaij:2011jh} a smaller data sample is used to determine
the non-prompt \Jpsi\ fractions in bins of \pTJpsi\ and rapidity.
These results are applied to the present \Jpsi\ sample without repeating
the full analysis.

\section{Efficiencies}
\label{sec:Efficiencies}

\begin{figure}[!htb]
  \begin{center}
  \subfigure{
    \ifthenelse{\boolean{pdflatex}}{
      \includegraphics*[width=0.47\textwidth]{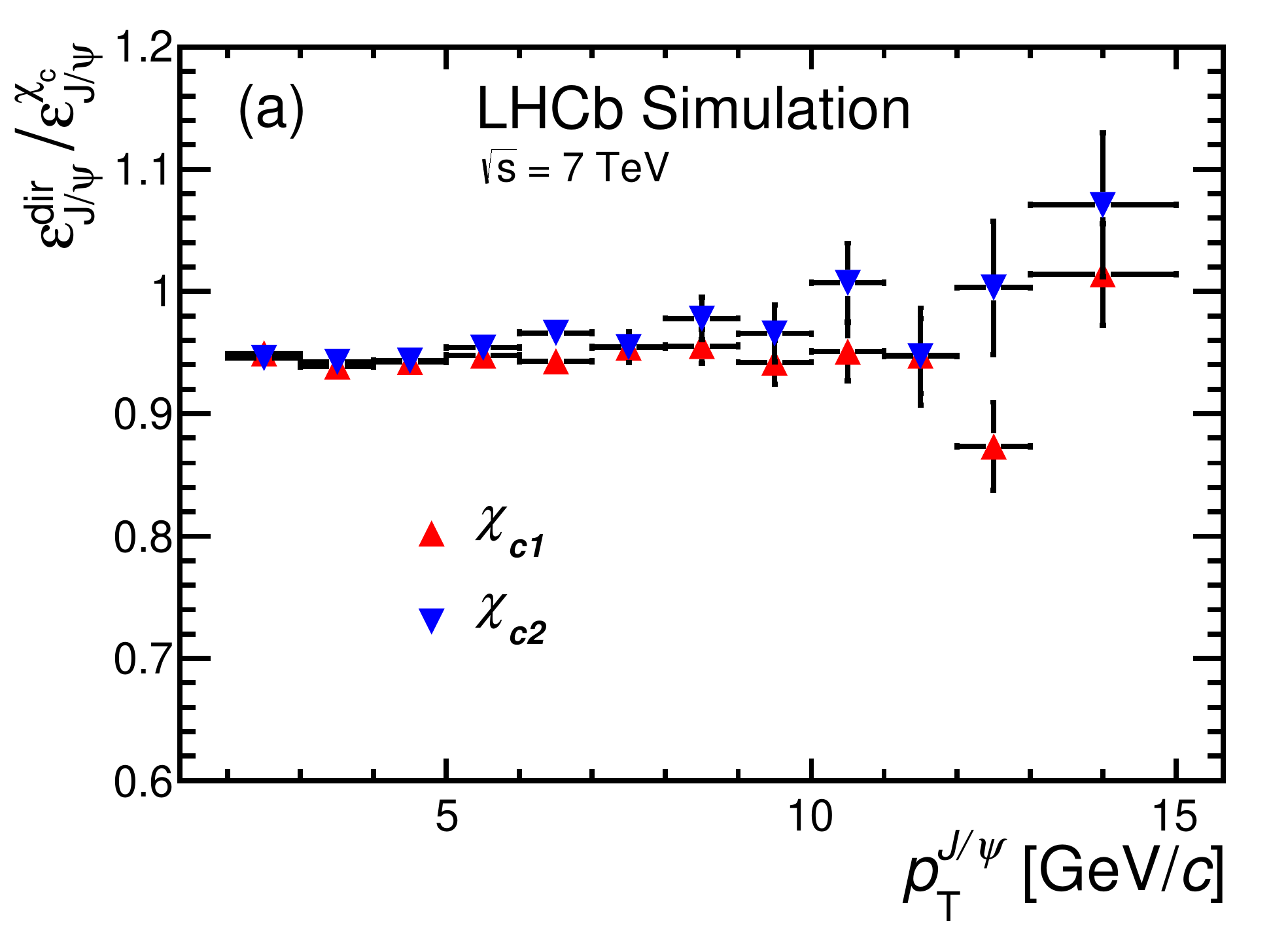}
    }{
      \includegraphics*[width=0.47\textwidth]{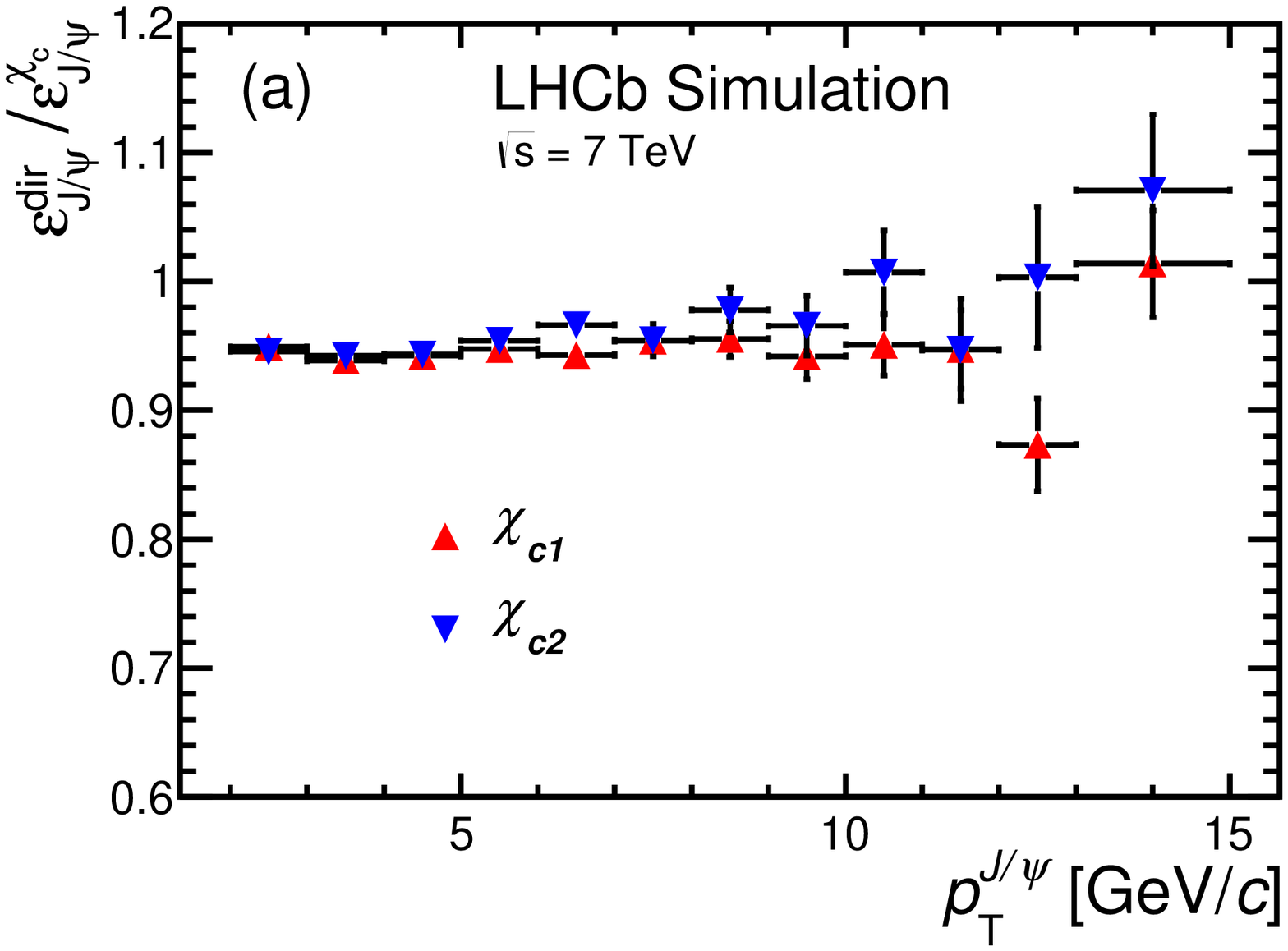}}}
      \subfigure{
    \ifthenelse{\boolean{pdflatex}}{
      \includegraphics*[width=0.47\textwidth]{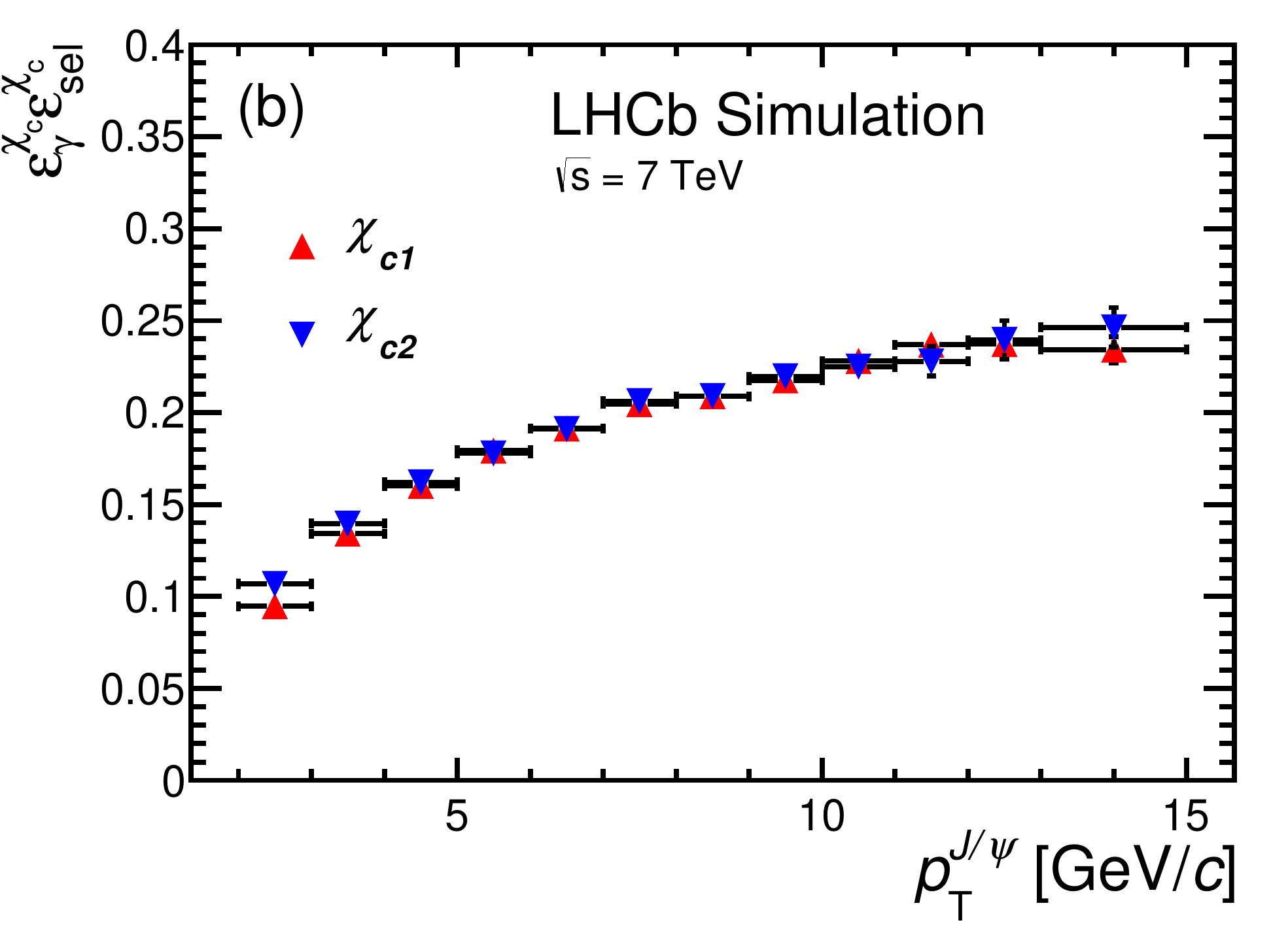}
    }{
      \includegraphics*[width=0.47\textwidth]{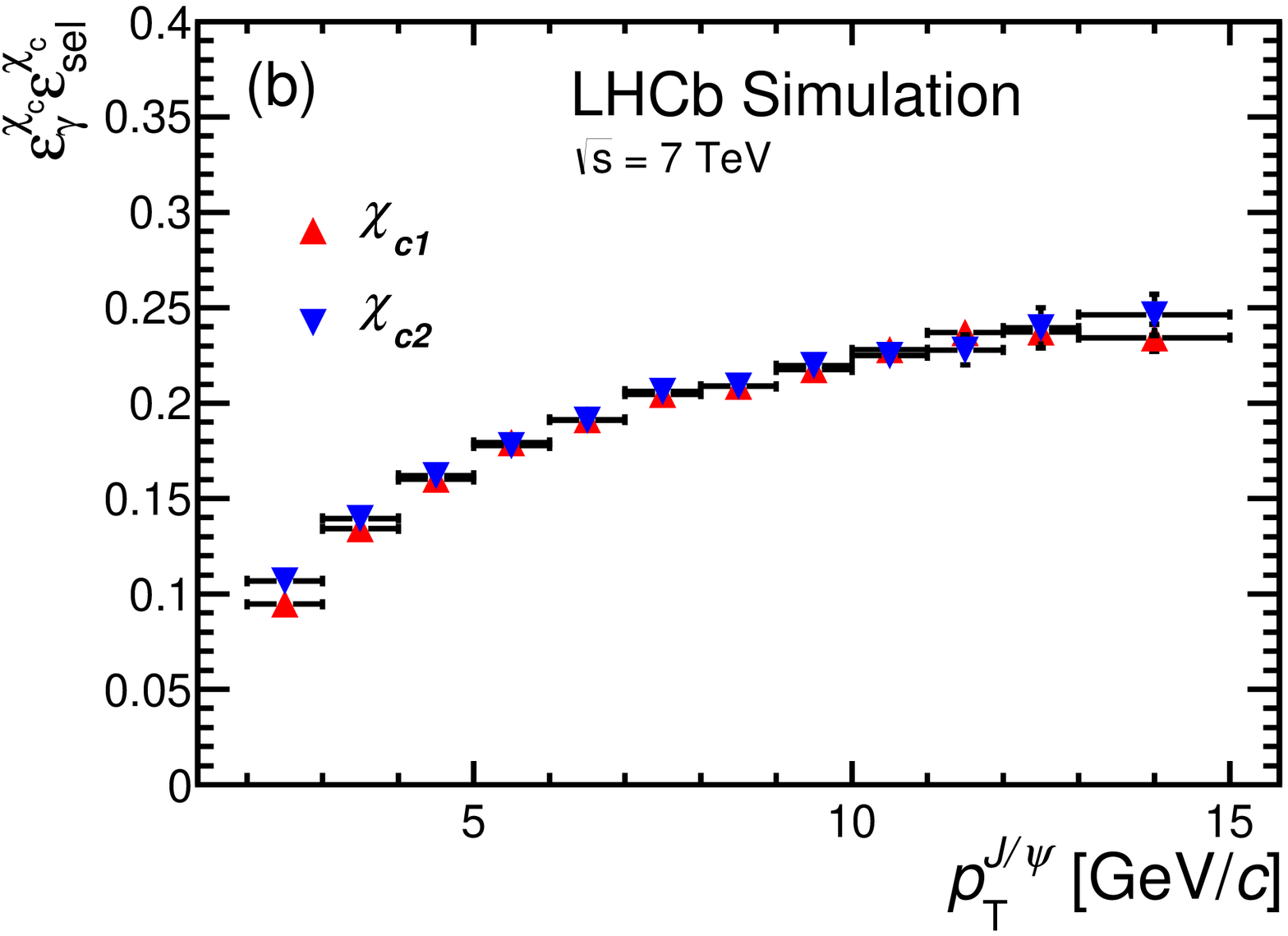}}}
  \caption{
    \label{fig:RecEff}
    \small{(a) Ratio of the reconstruction and selection efficiency for direct
    \Jpsi\ compared to \Jpsi\ from \Chic\ decays, 
    \ensuremath{\epsilon^{\mathrm{dir}}_{\Jpsi}/\epsilon^{\Chic}_{\Jpsi}}, 
    and (b) the photon reconstruction and selection efficiency 
    multiplied by the \Chic\ selection efficiency,
    \ensuremath{\epsilon^{\ChicJ}_{\gamma}\epsilon^{\ChicJ}_{\mathrm{sel}}},
    obtained from simulation.
    The efficiencies are presented separately for the
    \ChicOne\ (red triangles) and \ChicTwo\ (inverted blue triangles) states,
    and as a function of \pTJpsi.}}
	\end{center}
\end{figure}

The efficiencies to reconstruct and select \Jpsi\ and \Chic\ 
candidates are taken from simulation.
The efficiency ratio 
\ensuremath{\epsilon^{2S}_{\Jpsi}/\epsilon^{\mathrm{dir}}_{\Jpsi}}
is consistent with unity for all \pTJpsi\ bins;
hence, 
\ensuremath{R_{2S}} is set 
equal to 1 in \Equation~(\ref{e:ChiC_Jpsi_ratioR}).
The ratio of efficiencies
\ensuremath{\epsilon^{\mathrm{dir}}_{\Jpsi}/\epsilon^{\ChicJ}_{\Jpsi}} 
and the product of efficiencies 
\ensuremath{\epsilon^{\ChicJ}_{\gamma}\epsilon^{\ChicJ}_{\mathrm{sel}}}
for the \ChicOne\ and \ChicTwo\ states are shown in \Figure~\ref{fig:RecEff}. 
In general these efficiencies are the same for the two states, 
except at low \pTJpsi\ where the
reconstruction and detection efficiencies for 
\ChicTwo\ are significantly larger than for \ChicOne. 
This difference arises from the effect of the requirement
\ensuremath{\pTGamma\myop{>}\myvalue{650}{\MeVc}} 
which results in more photons surviving from \ChicTwo\ decays than 
from \ChicOne\ decays. 

The photon detection efficiency obtained using simulation 
is validated using candidate
\BToJpsiK\ and \BToChicK\ (including charge conjugate) decays 
selected from the same data set as the prompt \Jpsi\ and \Chic\ candidates.
The efficiency to reconstruct and select a photon
from a \Chic\ in \BToChicK\ decays,
\ensuremath{\epsilon_{\gamma}},
is evaluated using 
\begin{align}
\label{e:BToChicK_BToJpsiK_ratio}
\epsilon_{\gamma}
\myop{=}
\frac{N_{\BToChicK}}{N_{\BToJpsiK}}
\times \frac{\brBToJpsiK}{\brBToChicK \cdot \brChicToJpsiGamma} 
\times R_{\epsilon} 
\end{align}
where \ensuremath{N_{\BToChicK}} and \ensuremath{N_{\BToJpsiK}} are the
measured yields of \BToChicK\ and \BToJpsiK\ and
\ensuremath{\BR} are the known branching fractions.
The factor \ensuremath{R_{\epsilon}=1.04\pm 0.02}
is obtained from simulation and takes into account
any differences in the acceptance,
trigger,
selection and reconstruction efficiencies of the
\ensuremath{K}, \Jpsi, \Chic\ (except the photon detection efficiency) 
and \ensuremath{B^+} in \BToJpsiK\ and \BToChicK\ decays.
All branching fractions are taken from 
\Reference~\cite{Nakamura:2010zzi}.
The \BToJpsiK\ branching fraction is 
\ensuremath{\brBToJpsiK = \brBToJpsiKmeas}.   
The dominant process for 
\ensuremath{\BToChicK\rightarrow\Jpsi\gamma K^+} 
decays is via the \ChicOne\ state,  
with branching fractions 
\ensuremath{\brBToChicOneK = \brBToChicOneKmeas} and
\ensuremath{\brChicOneToJpsiGamma = \brChicOneToJpsiGammameas};
the contributions from the \ChicZero\ and \ChicTwo\ modes 
are neglected.
  
The \BToChicK\ and \BToJpsiK\ candidates are selected keeping as many of the
selection criteria in common as possible with the main analysis. 
The \Jpsi\ and \Chic\ selection 
criteria are the same as for the prompt analysis,
apart from the pseudo-decay time requirement.
The bachelor kaon is required to have 
a well measured track 
(\ensuremath{\chi^2/\mathrm{ndf}<5}), 
a minimum impact parameter \ensuremath{\chi^2 } 
with respect to all primary vertices of
greater than \ensuremath{9} 
and
a momentum greater than \ensuremath{5\, \GeVc}.
The bachelor is identified as a kaon by the RICH detectors 
by requiring the difference in log-likelihoods between
the kaon and pion hypotheses to be larger than \ensuremath{5}.
The \ensuremath{B} candidate is formed from the \Chic\ or \Jpsi\ candidate 
and the bachelor kaon.
The \ensuremath{B} vertex is required to be 
well measured (\ensuremath{\chi^2/\mathrm{ndf}<9})
and separated from the primary vertex 
(flight distance \ensuremath{\chi^2 >50}).
The \ensuremath{B} momentum vector is required to point towards 
the primary vertex
(\ensuremath{\cos\theta>0.9999}, 
where \ensuremath{\theta} is the angle between the \ensuremath{B} momentum and
the direction between the primary and \ensuremath{B} vertices)   
and have an impact parameter \ensuremath{\chi^2} smaller than \ensuremath{9}.
The combinatorial background under the \Chic\ peak for the \BToChicK\
candidates is reduced by requiring the mass difference
\ensuremath{\DeltaM_{\Chic}= M(\mup\mun\gamma)-M(\mup\mun)< 600\MeVcc}.
A small number of \BToChicK\ candidates 
which form a good \BToJpsiK\ candidate are
removed by requiring
\ensuremath{|M(\mup\mun\gamma K)-M(\mup\mun K)|>200\MeVcc}. 

\begin{figure}[!htb]
  \begin{center}
    \subfigure{
      \ifthenelse{\boolean{pdflatex}}{
	\includegraphics*[width=0.45\textwidth]{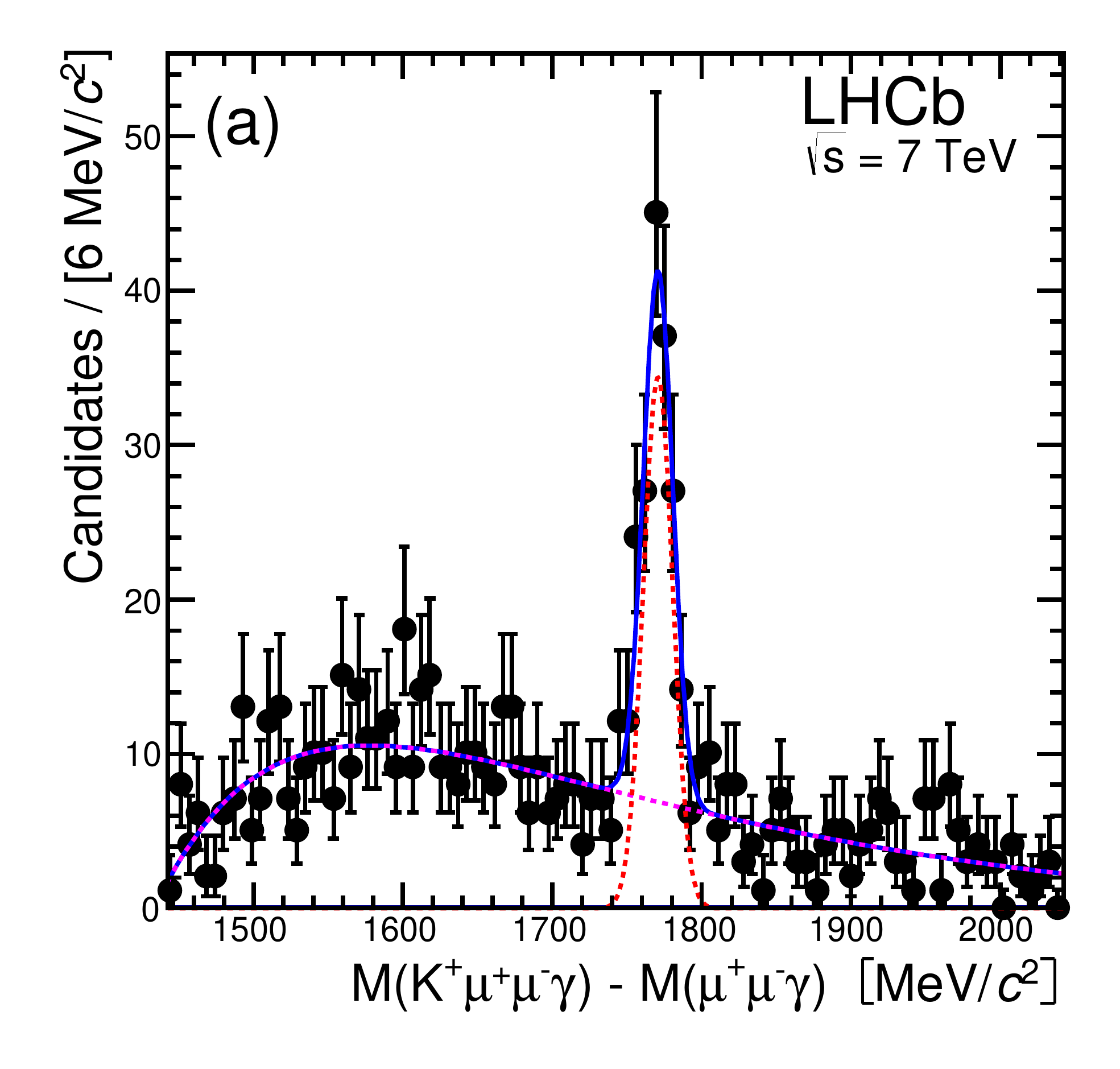}
      }{
	\includegraphics*[width=0.45\textwidth]{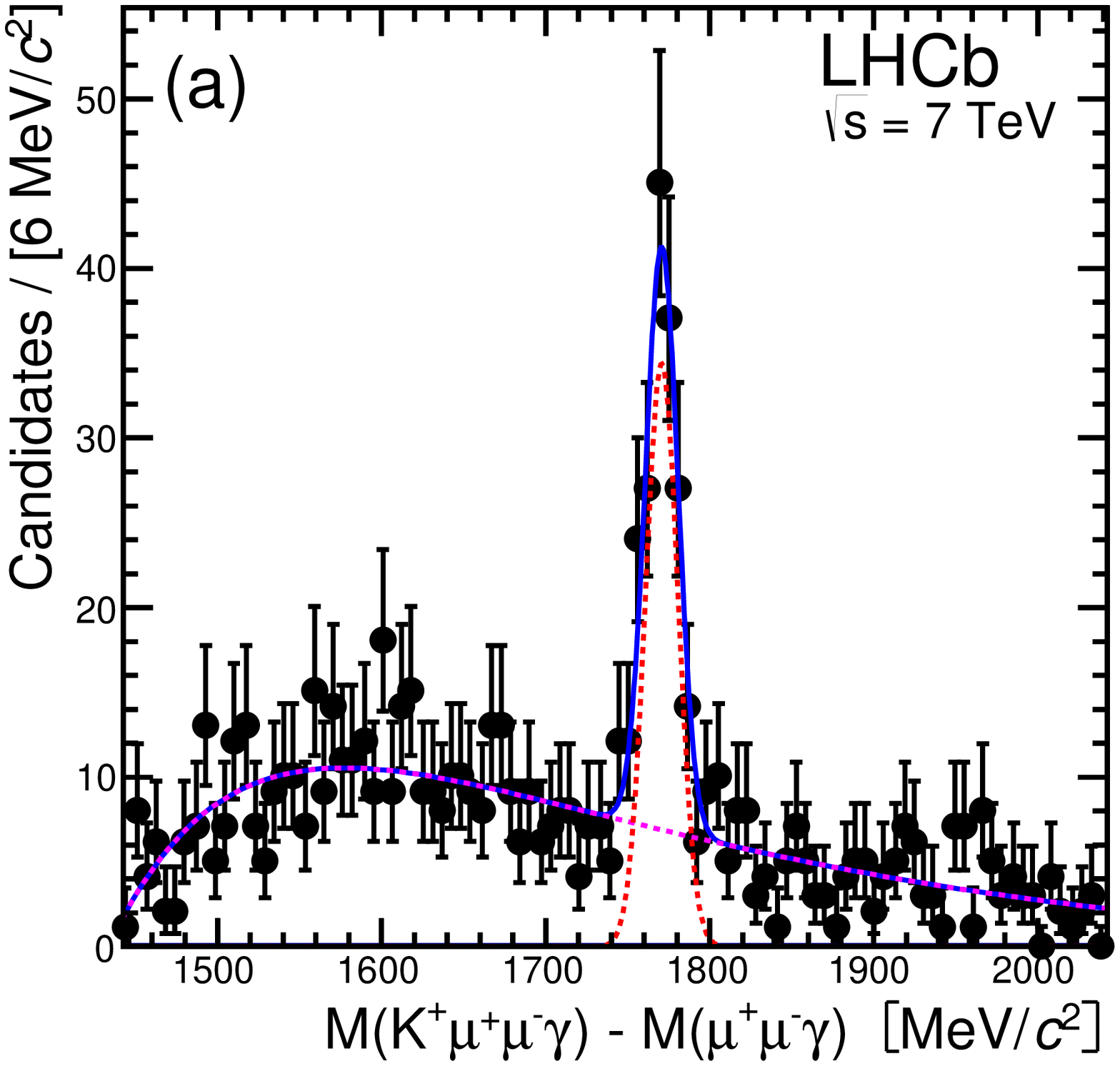}}}
    \subfigure{
      \ifthenelse{\boolean{pdflatex}}{
	\includegraphics*[width=0.45\textwidth]{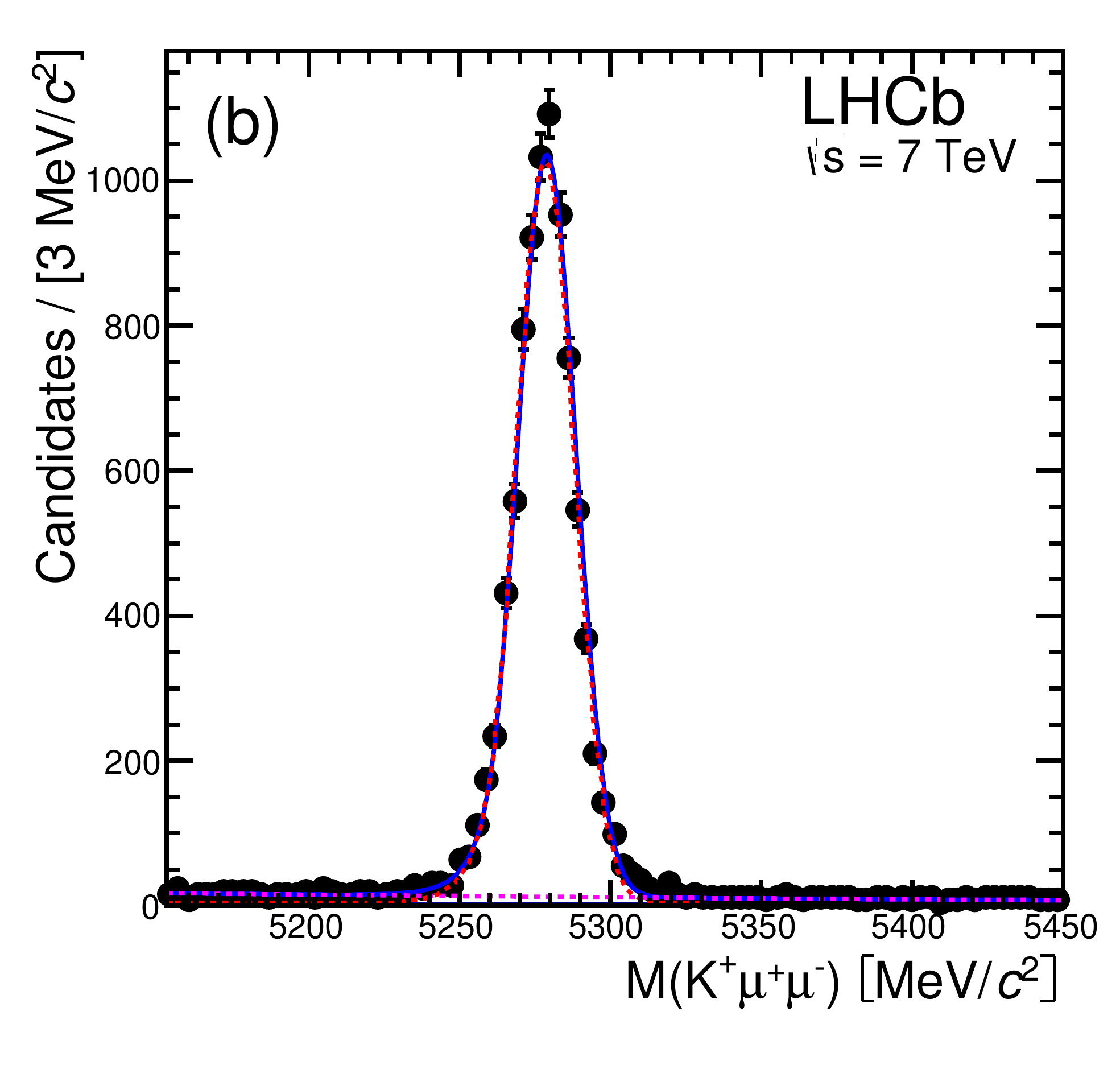}
      }{
	\includegraphics*[width=0.45\textwidth]{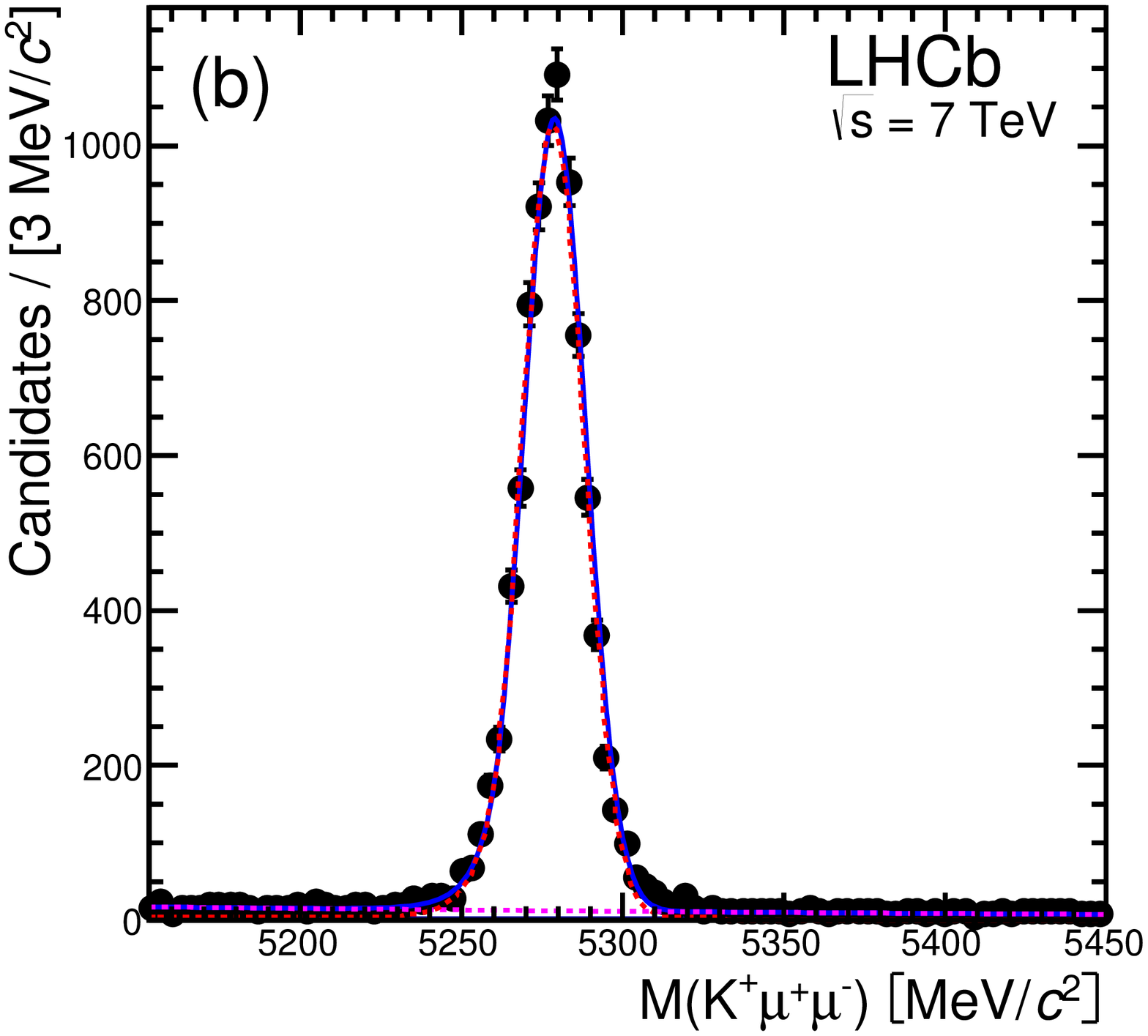}}}
    \caption{\small{(a) Reconstructed 
	\ensuremath{\Delta M_{B^+}=M(\mup\mun\gamma K)-M(\mup\mun\gamma)}
	mass distribution for \BToChicK\ candidates and
	(b) the reconstructed \ensuremath{B^+} mass distribution for
	\BToJpsiK\ candidates. 
	The LHCb data are shown as solid black points,
	the full fit functions with a solid blue (upper) curve,
	the contribution from signal candidates with a dashed red 
	(lower curve) and
	the background with a dashed purple curve.}}
    \label{fig:BToChicK}
  \end{center}
\end{figure}

The \ensuremath{\Delta M_{B^+}=M(\mup\mun\gamma K)-M(\mup\mun\gamma)}
mass distribution for the \BToChicK\ candidates   
is shown in \Figure~\ref{fig:BToChicK}(a);
\ensuremath{\Delta M_{B^+}} is computed to improve the
resolution and hence the signal-to-background ratio.
The \BToChicK\ yield, 
\yBToChicKmeas\ candidates, 
is determined from a fit that uses a Gaussian function
to describe the signal peak and a threshold function, 
\begin{align}
  f(x)\myop{=}x^{a}\!\left(1-e^{\frac{m_{0}}{c}\left(1-x\right)}\right)+b\left(x-1\right),
\end{align}
where 
\ensuremath{x\myop{=}\Delta M_{B^+}\myop{/}m_{0}} 
and $m_0$, $a$, $b$ and $c$ are free parameters,
to model the background.
The reconstructed \ensuremath{B^+} mass distribution for
the \BToJpsiK\ candidates is shown in \Figure~\ref{fig:BToChicK}(b).
The \BToJpsiK\ yield, \yBToJpsiKmeas\ candidates, 
is determined from a fit 
that uses a Crystal Ball  
function~\cite{Skwarnicki:1986xj} 
to describe the signal peak 
and an exponential to model the background.

The photon efficiency from the observation of \BToChicK\ and \BToJpsiK\
decays is measured to be 
\ensuremath{\epsilon_{\gamma}=(11.3\pm 1.2\pm 1.2)\%}
where the first error is statistical 
and is dominated by the observed yield of \BToChicK\
candidates,
and the second error is systematic and is given 
by the uncertainty on the branching fraction \brBToChicOneK. 
The photon efficiency measured in data can be
compared to the photon efficiency,
\ensuremath{(11.7\pm 0.3)\%},
obtained using the same procedure on simulated events.
The measurements are in good agreement
and the uncertainty on the difference between data and simulation
is propagated as a
\ensuremath{\pm 14\%} relative systematic uncertainty 
on the photon efficiency in the measurement of
\ensuremath{\srChicToJpsi}.
%

\section{Polarisation}
\label{sec:Polarisation}

The simulation used to calculate the efficiencies 
and, hence, extract the result of \Equation~(\ref{e:ChiC_Jpsi_ratio})
assumes that the \Jpsi\ and \Chic\ are unpolarised.
The effect of polarised states is studied by reweighting the 
simulated events according to different polarisation scenarios;
the results are shown in \Table~\ref{tab:PolFactors}.
It is also noted that, 
since the \PsiTwoS\ decays predominantly to \ensuremath{\Jpsi\pi\pi}, 
with the \ensuremath{\pi\pi} in an $S$ wave state~\cite{bai:1999mj},
and the \PsiTwoS\ polarisation should not differ significantly
from the polarisation of directly produced \Jpsi\ mesons,
the effect of the polarisation can be considered 
independent of the \PsiTwoSToJpsiX\ contribution~\cite{Faccioli:2012kp}. 

The \Jpsi\ and \ChicToJpsiGamma\ angular distributions are calculated in
the helicity frame assuming azimuthal symmetry.
This choice of reference frame provides an estimate of the effect
of polarisation on the results,
pending the direct measurements of the \Jpsi\ and \Chic\ polarisations.
The \Jpsi\ system
is described by the angle 
\ensuremath{\theta_{\Jpsi}},
which is the angle between the directions of the 
\ensuremath{\mup} 
in the \Jpsi\ rest frame and the 
\Jpsi\ in the laboratory frame.
The \ensuremath{\theta_{\Jpsi}} distribution
depends on the parameter
\ensuremath{\lambda_{\Jpsi}} which describes the \Jpsi\ polarisation;
\ensuremath{\lambda_{\Jpsi}=+1, -1, 0} corresponds to
pure transverse, pure longitudinal and no polarisation respectively.
The \ChicToJpsiGamma\ system
is described by 
three angles: 
\ensuremath{\theta^{\prime}_{\Jpsi}}, 
\ensuremath{\theta_{\Chic}} and \ensuremath{\phi},
where 
\ensuremath{\theta^{\prime}_{\Jpsi}} is the angle between the directions
of the \ensuremath{\mup} in the \Jpsi\ rest frame and the \Jpsi\ in the \Chic\ rest frame,
\ensuremath{\theta_{\Chic}} is the angle between the
directions of the \Jpsi\ in the \Chic\ rest frame and the \Chic\ in the
laboratory frame,
and 
\ensuremath{\phi} is the angle between 
the \Jpsi\ decay plane in the \Chic\ rest frame and the 
plane formed by
the \Chic\ direction in the laboratory frame and the direction of the \Jpsi\ in the \Chic\ rest frame. 
The general expressions for the
angular distributions are independent of the choice of polarisation axis 
(here chosen as the direction of the \Chic\ in the laboratory frame)
and are detailed in \Reference~\cite{Abt:2008ed}. 
The angular distributions of the \Chic\ states depend on 
\ensuremath{m_{\chi_{cJ}}} which is the 
azimuthal angular momentum quantum number
of the 
\ensuremath{\chi_{cJ}} state.  

For each simulated event in the unpolarised sample, 
a weight is calculated from the distributions of
\ensuremath{\theta^{\prime}_{\Jpsi}},
\ensuremath{\theta_{\Chic}} and \ensuremath{\phi}
in the various polarisation hypotheses compared to the 
unpolarised distributions.
The weights shown in \Table~\ref{tab:PolFactors} are then the average
of these per-event weights in the simulated sample.
For a given 
(\ensuremath{|m_{\ChicOne}|}, \ensuremath{|m_{\ChicTwo}|},
\ensuremath{\lambda_{\Jpsi}})
polarisation combination, the central value 
of the determined cross-section ratio 
in each \pTJpsi\ bin should be multiplied by the
number in the table.
The maximum effect from the possible polarisation of 
the \Jpsi, \ChicOne\ and \ChicTwo\ mesons is
given separately from the systematic uncertainties in 
Table~\ref{tab:ResultsEqOne} and \Figure~\ref{fig:Results}.
\begin{table*}[htbp]
\tiny
\renewcommand{\arraystretch}{1.8}
\setlength{\tabcolsep}{1mm}
\caption{
  \label{tab:PolFactors}
  \small{Polarisation weights in \pTJpsi\ bins for different combinations of the
    \Jpsi, \ChicOne\ and \ChicTwo\ polarisations.
    \ensuremath{\lambda_{\Jpsi}} is the \Jpsi\ polarisation parameter; 
    \ensuremath{\lambda_{\Jpsi}=+1, -1, 0} corresponds to
    fully transverse, fully longitudinal and no polarisation respectively. 
    \ensuremath{m_{\ChicJ}} is the 
    azimuthal angular momentum quantum number 
    corresponding to  
    total angular momentum $J$; Unpol means the \Chic\ is unpolarised.}}
\smallskip
\begin{center} 
\resizebox{12cm}{!}{
\begin{tabular}{|c|c|c|c|c|c|c|c|c|c|c|c|c|c|} \hline
\multirow{2}{*}{(\ensuremath{|m_{\ChicOne}|,|m_{\ChicTwo}|,\lambda_{\Jpsi}})} & \multicolumn{12}{|c|}{\pTJpsi\ (\GeVc)} \\ \cline{2-13}
 & 2-3 & 3-4 & 4-5 & 5-6 & 6-7 & 7-8 & 8-9 & 9-10 & 10-11 & 11-12 & 12-13 & 13-15\\ \hline
(Unpol,Unpol,-1)  & 1.16 & 1.15 & 1.15 & 1.15 & 1.15 & 1.14 & 1.14 & 1.13 & 1.12 & 1.12 & 1.10 & 1.10 \\ 
(Unpol,Unpol,1)  & 0.92 & 0.92 & 0.92 & 0.92 & 0.92 & 0.92 & 0.93 & 0.93 & 0.93 & 0.94 & 0.95 & 0.94 \\ 
(Unpol,0,-1)  & 1.16 & 1.14 & 1.13 & 1.11 & 1.10 & 1.09 & 1.09 & 1.08 & 1.07 & 1.06 & 1.06 & 1.07 \\ 
(Unpol,0,0)  & 1.00 & 0.99 & 0.98 & 0.97 & 0.96 & 0.95 & 0.95 & 0.96 & 0.95 & 0.95 & 0.96 & 0.97 \\ 
(Unpol,0,1)  & 0.91 & 0.91 & 0.90 & 0.89 & 0.88 & 0.87 & 0.88 & 0.89 & 0.89 & 0.88 & 0.91 & 0.92 \\ 
(Unpol,1,-1)  & 1.15 & 1.14 & 1.14 & 1.13 & 1.13 & 1.12 & 1.11 & 1.11 & 1.10 & 1.09 & 1.08 & 1.09 \\ 
(Unpol,1,0)  & 0.99 & 0.99 & 0.99 & 0.98 & 0.98 & 0.98 & 0.98 & 0.98 & 0.98 & 0.98 & 0.98 & 0.98 \\ 
(Unpol,1,1)  & 0.90 & 0.91 & 0.91 & 0.90 & 0.90 & 0.90 & 0.91 & 0.91 & 0.91 & 0.91 & 0.93 & 0.93 \\ 
(Unpol,2,-1)  & 1.18 & 1.17 & 1.18 & 1.20 & 1.21 & 1.21 & 1.20 & 1.19 & 1.19 & 1.19 & 1.16 & 1.15 \\ 
(Unpol,2,0)  & 1.01 & 1.02 & 1.03 & 1.04 & 1.05 & 1.06 & 1.06 & 1.05 & 1.06 & 1.07 & 1.05 & 1.04 \\ 
(Unpol,2,1)  & 0.93 & 0.94 & 0.94 & 0.96 & 0.97 & 0.98 & 0.98 & 0.98 & 0.99 & 1.00 & 1.00 & 0.99 \\ 
(0,Unpol,-1)  & 1.16 & 1.15 & 1.18 & 1.21 & 1.22 & 1.23 & 1.25 & 1.25 & 1.26 & 1.22 & 1.23 & 1.25 \\ 
(0,Unpol,0) & 0.99 & 1.00 & 1.02 & 1.05 & 1.07 & 1.08 & 1.10 & 1.11 & 1.12 & 1.10 & 1.12 & 1.14 \\ 
(0,Unpol,1)  & 0.91 & 0.93 & 0.94 & 0.97 & 0.98 & 1.00 & 1.02 & 1.04 & 1.05 & 1.03 & 1.06 & 1.08 \\ 
(1,Unpol,-1)  & 1.17 & 1.15 & 1.14 & 1.13 & 1.12 & 1.11 & 1.09 & 1.08 & 1.07 & 1.08 & 1.05 & 1.05 \\ 
(1,Unpol,0) & 1.00 & 1.00 & 0.99 & 0.98 & 0.97 & 0.97 & 0.96 & 0.95 & 0.95 & 0.96 & 0.95 & 0.95 \\ 
(1,Unpol,1)  & 0.92 & 0.92 & 0.91 & 0.90 & 0.89 & 0.89 & 0.89 & 0.89 & 0.89 & 0.90 & 0.90 & 0.89 \\ 
(0,0,-1)  & 1.15 & 1.14 & 1.15 & 1.17 & 1.18 & 1.18 & 1.20 & 1.21 & 1.20 & 1.17 & 1.19 & 1.22 \\ 
(0,0,0)  & 0.99 & 0.99 & 1.00 & 1.02 & 1.02 & 1.03 & 1.05 & 1.07 & 1.07 & 1.04 & 1.08 & 1.11 \\ 
(0,0,1)  & 0.91 & 0.91 & 0.92 & 0.93 & 0.94 & 0.95 & 0.98 & 1.00 & 1.00 & 0.98 & 1.02 & 1.05 \\ 
(0,1,-1)  & 1.14 & 1.14 & 1.16 & 1.19 & 1.20 & 1.21 & 1.22 & 1.23 & 1.23 & 1.20 & 1.21 & 1.24 \\ 
(0,1,0)  & 0.98 & 0.99 & 1.01 & 1.03 & 1.05 & 1.06 & 1.08 & 1.09 & 1.10 & 1.07 & 1.10 & 1.12 \\ 
(0,1,1)  & 0.90 & 0.92 & 0.93 & 0.95 & 0.96 & 0.98 & 1.00 & 1.02 & 1.03 & 1.01 & 1.04 & 1.07 \\ 
(0,2,-1)  & 1.17 & 1.17 & 1.21 & 1.25 & 1.29 & 1.30 & 1.31 & 1.31 & 1.32 & 1.30 & 1.28 & 1.30 \\ 
(0,2,0)  & 1.01 & 1.02 & 1.05 & 1.09 & 1.12 & 1.14 & 1.16 & 1.17 & 1.19 & 1.17 & 1.17 & 1.18 \\ 
(0,2,1)  & 0.92 & 0.94 & 0.96 & 1.01 & 1.03 & 1.06 & 1.08 & 1.09 & 1.11 & 1.10 & 1.11 & 1.12 \\ 
(1,0,-1)  & 1.16 & 1.13 & 1.12 & 1.09 & 1.07 & 1.05 & 1.04 & 1.04 & 1.02 & 1.02 & 1.01 & 1.01 \\ 
(1,0,0)  & 1.00 & 0.99 & 0.97 & 0.94 & 0.93 & 0.92 & 0.91 & 0.91 & 0.90 & 0.91 & 0.91 & 0.92 \\ 
(1,0,1)  & 0.92 & 0.91 & 0.89 & 0.87 & 0.85 & 0.84 & 0.85 & 0.85 & 0.84 & 0.85 & 0.86 & 0.86 \\ 
(1,1,-1)  & 1.15 & 1.14 & 1.13 & 1.11 & 1.10 & 1.08 & 1.07 & 1.06 & 1.05 & 1.05 & 1.03 & 1.03 \\ 
(1,1,0)  & 0.99 & 0.99 & 0.98 & 0.96 & 0.95 & 0.94 & 0.94 & 0.94 & 0.93 & 0.94 & 0.94 & 0.93 \\ 
(1,1,1)  & 0.91 & 0.91 & 0.90 & 0.88 & 0.87 & 0.87 & 0.87 & 0.87 & 0.87 & 0.88 & 0.88 & 0.88 \\ 
(1,2,-1)  & 1.18 & 1.17 & 1.17 & 1.17 & 1.18 & 1.18 & 1.16 & 1.14 & 1.14 & 1.15 & 1.11 & 1.09 \\ 
(1,2,0)  & 1.02 & 1.01 & 1.01 & 1.02 & 1.03 & 1.03 & 1.02 & 1.01 & 1.02 & 1.03 & 1.01 & 0.99 \\ 
(1,2,1)  & 0.93 & 0.94 & 0.93 & 0.94 & 0.94 & 0.95 & 0.94 & 0.94 & 0.95 & 0.97 & 0.95 & 0.93\\ \hline
\end{tabular}
}
\end{center}
\end{table*}

\section{Systematic uncertainties}
\label{sec:Systematics}

\afterpage{\clearpage}

\begin{table*}[htbp] 
\footnotesize
\renewcommand{\arraystretch}{2}
\caption{
  \label{tab:Systematics1}
  \small{Summary of the systematic uncertainties on \srChicToJpsi\ in each \pTJpsi\ bin.}}
\smallskip
\begin{center}
\begin{tabular}{|c|c|c|c|c|c|c|} \hline
\pTJpsi (\GeVc)    & $2-3$           & $3-4$          & $4-5$     & $5-6$   & $6-7$    & $7-8$      \\ \hline
Size of simulation sample  & ${}^{+0.0006}_{-0.0005}$ & ${}^{+0.0006}_{-0.0005}$ & ${}^{+0.0007}_{-0.0006}$ & ${}^{+0.0009}_{-0.0009}$ & ${}^{+0.001}_{-0.001}$ & ${}^{+0.002}_{-0.002}$ \\
Photon efficiency  & ${}^{+0.011}_{-0.010}$ & ${}^{+0.013}_{-0.011}$ & ${}^{+0.013}_{-0.012}$ & ${}^{+0.016}_{-0.013}$ & ${}^{+0.016}_{-0.013}$ & ${}^{+0.017}_{-0.015}$ \\
Non-prompt \Jpsi\ fraction  & ${}^{+0.002}_{-0.005}$ & ${}^{+0.003}_{-0.005}$ & ${}^{+0.003}_{-0.006}$ & ${}^{+0.004}_{-0.008}$ & ${}^{+0.005}_{-0.010}$ & ${}^{+0.006}_{-0.011}$ \\
Fit model  & ${}^{+0.003}_{-0.003}$ & ${}^{+0.003}_{-0.003}$ & ${}^{+0.002}_{-0.004}$ & ${}^{+0.003}_{-0.005}$ & ${}^{+0.002}_{-0.005}$ & ${}^{+0.003}_{-0.006}$ \\
Simulation calibration & ${}^{+0.010}_{-0.000}$ & ${}^{+0.010}_{-0.000}$ & ${}^{+0.012}_{-0.000}$ & ${}^{+0.012}_{-0.000}$ & ${}^{+0.015}_{-0.000}$ & ${}^{+0.014}_{-0.000}$ \\
 \hline
\pTJpsi (\GeVc)        & $8-9$    & $9-10$  & $10-11$     & $11-12$   & $12-13$    & $13-15$   \\ \hline
Size of simulation sample  & ${}^{+0.002}_{-0.002}$ & ${}^{+0.003}_{-0.003}$ & ${}^{+0.004}_{-0.004}$ & ${}^{+0.006}_{-0.006}$ & ${}^{+0.008}_{-0.008}$ & ${}^{+0.008}_{-0.008}$ \\
Photon efficiency & ${}^{+0.018}_{-0.016}$ & ${}^{+0.020}_{-0.016}$ & ${}^{+0.019}_{-0.016}$ & ${}^{+0.019}_{-0.018}$ & ${}^{+0.021}_{-0.020}$ & ${}^{+0.023}_{-0.019}$ \\
Non-prompt \Jpsi\ fraction & ${}^{+0.009}_{-0.011}$ & ${}^{+0.012}_{-0.013}$ & ${}^{+0.011}_{-0.017}$ & ${}^{+0.019}_{-0.019}$ & ${}^{+0.022}_{-0.018}$ & ${}^{+0.018}_{-0.010}$ \\
Fit model     & ${}^{+0.002}_{-0.005}$ & ${}^{+0.002}_{-0.003}$ & ${}^{+0.006}_{-0.002}$ & ${}^{+0.001}_{-0.006}$ & ${}^{+0.003}_{-0.008}$ & ${}^{+0.002}_{-0.004}$ \\
Simulation calibration  & ${}^{+0.015}_{-0.000}$ & ${}^{+0.017}_{-0.000}$ & ${}^{+0.018}_{-0.000}$ & ${}^{+0.018}_{-0.000}$ & ${}^{+0.017}_{-0.000}$ & ${}^{+0.022}_{-0.000}$ \\
 \hline
\end{tabular}
\end{center}
\end{table*}

The systematic uncertainties detailed below are measured by repeatedly 
sampling from the distribution of the parameter under consideration.
For each sampled value, the cross-section ratio is calculated and 
the \ensuremath{68.3\%} probability 
interval is determined from the resulting distribution. 

The statistical errors from the finite number of simulated events used
for the calculation of the efficiencies are included 
as a systematic uncertainty in the
final results. 
The uncertainty is determined by sampling the efficiencies used in 
\Equation~\ref{e:ChiC_Jpsi_ratio} according to their errors.
The relative systematic uncertainty due to the limited 
size of the simulation sample
is found to be in the range 
\myrange{(0.3}{3.2)\%}  
and is given for each
\pTJpsi\ bin in \Table~\ref{tab:Systematics1}.

The efficiency extracted from 
the simulation sample for reconstructing and selecting a photon in 
\ChicToJpsiGamma\ decays has been validated 
using \BToChicK\ and \BToJpsiK\ decays observed
in the data, as described in \Section~\ref{sec:Efficiencies}.
The relative uncertainty between the photon efficiencies measured in the data
and simulation,
\ensuremath{\pm 14\%},
arises from the finite size of the observed \BToChicK\ yield
and the uncertainty on the known \BToChicOneK\ branching fraction,
and is taken to be the systematic error assigned to the photon efficiency
in the measurement of \srChicToJpsi. 
The relative systematic uncertainty on the cross-section ratio
used in \Equation~\ref{e:ChiC_Jpsi_ratio}
is determined by sampling the photon efficiency according to its systematic 
error. It is found to be  in the range 
\myrange{(6.4}{8.7)\%} 
and is given for each \pTJpsi\ bin in 
\Table~\ref{tab:Systematics1}. 

The \Jpsi\ yield used in \Equation~\ref{e:ChiC_Jpsi_ratio} is
corrected for the fraction of non-prompt \Jpsi, taken from
\Reference~\cite{Aaij:2011jh}.
For those \pTJpsi\ and rapidity bins used in this analysis 
and not covered by \Reference~\cite{Aaij:2011jh}
(\ensuremath{13\myop{<}\pTJpsi\myop{<}14}\GeVc and 
\ensuremath{3.5\myop{<}\RapidityJpsi\myop{<}4.5};
\ensuremath{11\myop{<}\pTJpsi\myop{<}13}\GeVc and 
\ensuremath{4\myop{<}\RapidityJpsi\myop{<}4.5}; and
\ensuremath{14\myop{<}\pTJpsi\myop{<}15}\GeVc),
a linear extrapolation is performed,
allowing for asymmetric errors.
The systematic uncertainty on the cross-section ratio
is determined by sampling the non-prompt \Jpsi\ fraction according to a 
bifurcated Gaussian function.
The relative systematic uncertainty from the non-prompt \Jpsi\ fraction
is found to be
in the range \myrange{(1.3}{10.7)\%}
and is given for each \pTJpsi\ bin in \Table~\ref{tab:Systematics1}.   

The method used to determine the systematic uncertainty due to 
the fit procedure in the extraction of the \Chic\ yields 
is discussed in detail in \Reference~\cite{Aaij:2011chic}.
The uncertainty includes contributions from
uncertainties on the fixed parameters,
the fit range 
and the shape of the overall fit function.
The overall relative systematic uncertainty from the fit is found 
to be in the range
\myrange{(0.4}{3.2)\%}  
and is given for each bin of \pTJpsi\ in
\Table~\ref{tab:Systematics1}.

The systematic uncertainty related to the 
calibration of the simulation sample 
is evaluated by performing the full analysis
using simulated events and 
comparing to the expected cross-section ratio from simulated signal events.
The results give an underestimate of \ensuremath{10.9\%}
in the measurement of the \srChicToJpsi\ cross-section ratio.
This deviation is caused by non-Gaussian
signal shapes in the simulation which
arise from an untuned calorimeter calibration.
These are not seen in the data, which is well described
by Gaussian signal shapes. 
This deviation is included as a systematic error by sampling
from the negative half of a Gaussian with zero mean and a 
width of \ensuremath{10.9\%}.
The relative uncertainty on the cross-section ratio is found to be in the 
range 
\myrange{(6.3}{8.2)\%}  
and is given for each bin of \pTJpsi\ in
\Table~\ref{tab:Systematics1}.
A second check of the procedure was performed using simulated events generated according to 
the distributions observed in the data,  
\ie\ three overlapping Gaussians and a background shape similar to that in
\Figure~\ref{fig:MassPlots}. 
In this case no evidence for a deviation was observed.
Other systematic uncertainties due to the modelling of the detector
in the simulation are negligible.

In summary, the overall systematic uncertainty is evaluated by simultaneously 
sampling the deviation of the cross-section ratio 
from the central value,
using the distributions of the cross-section ratios described above.
The systematic uncertainty is then determined from the
resulting distribution as described 
earlier in this section.
The separate systematic uncertainties are shown  
in bins of \pTJpsi\ in
\Table~\ref{tab:Systematics1}
and the combined uncertainties are shown in 
\Table~\ref{tab:ResultsEqOne}.

\section{Results and conclusions}
\label{sec:Results}

\begin{figure}[htbp]
  \begin{center}
    \subfigure{
    \ifthenelse{\boolean{pdflatex}}{
      \includegraphics*[width=8cm]{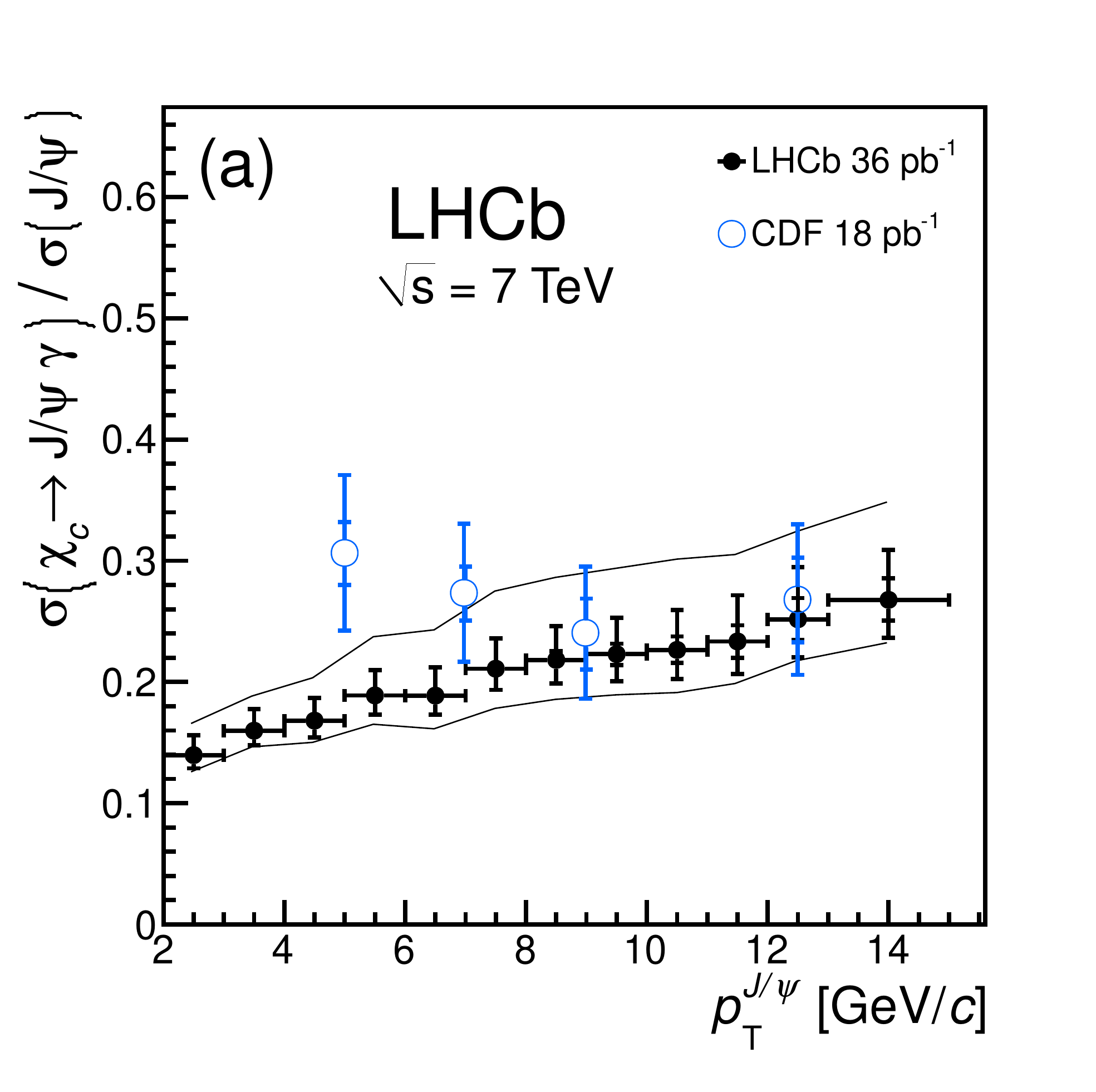}
    }{
      \includegraphics*[width=8cm]{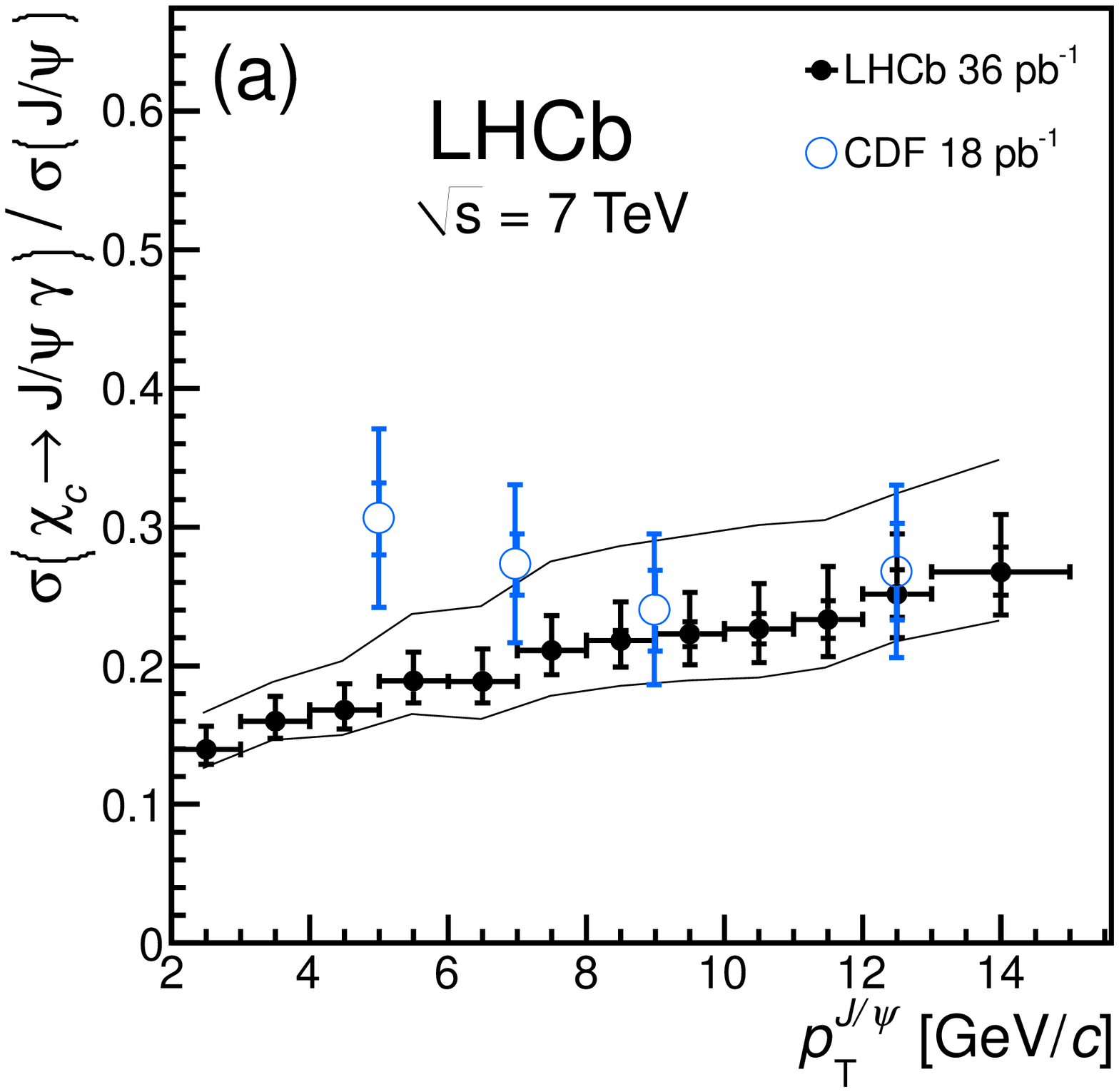}}}
    \subfigure{
    \ifthenelse{\boolean{pdflatex}}{
      \includegraphics*[width=8cm]{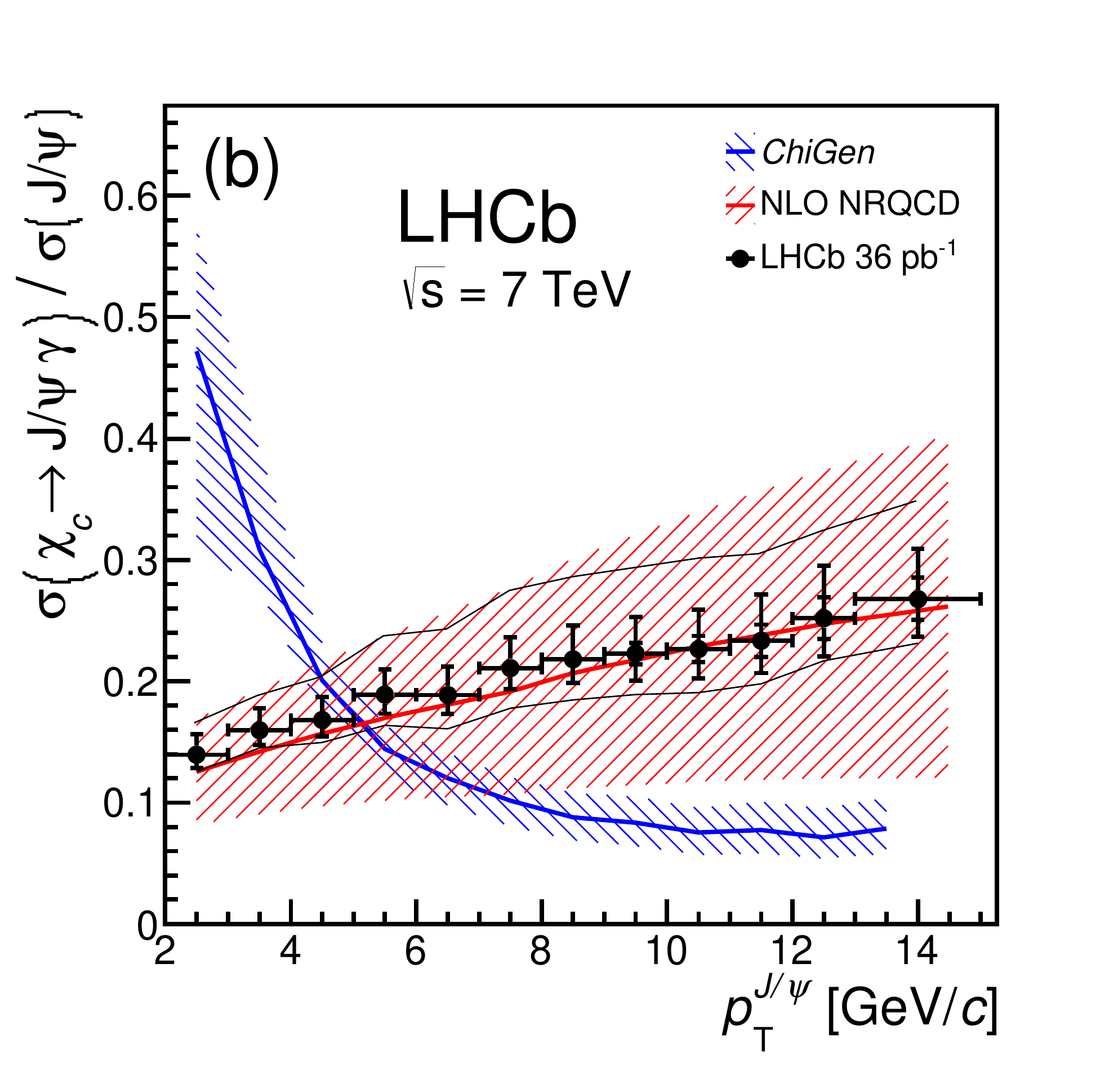}
    }{
      \includegraphics*[width=8cm]{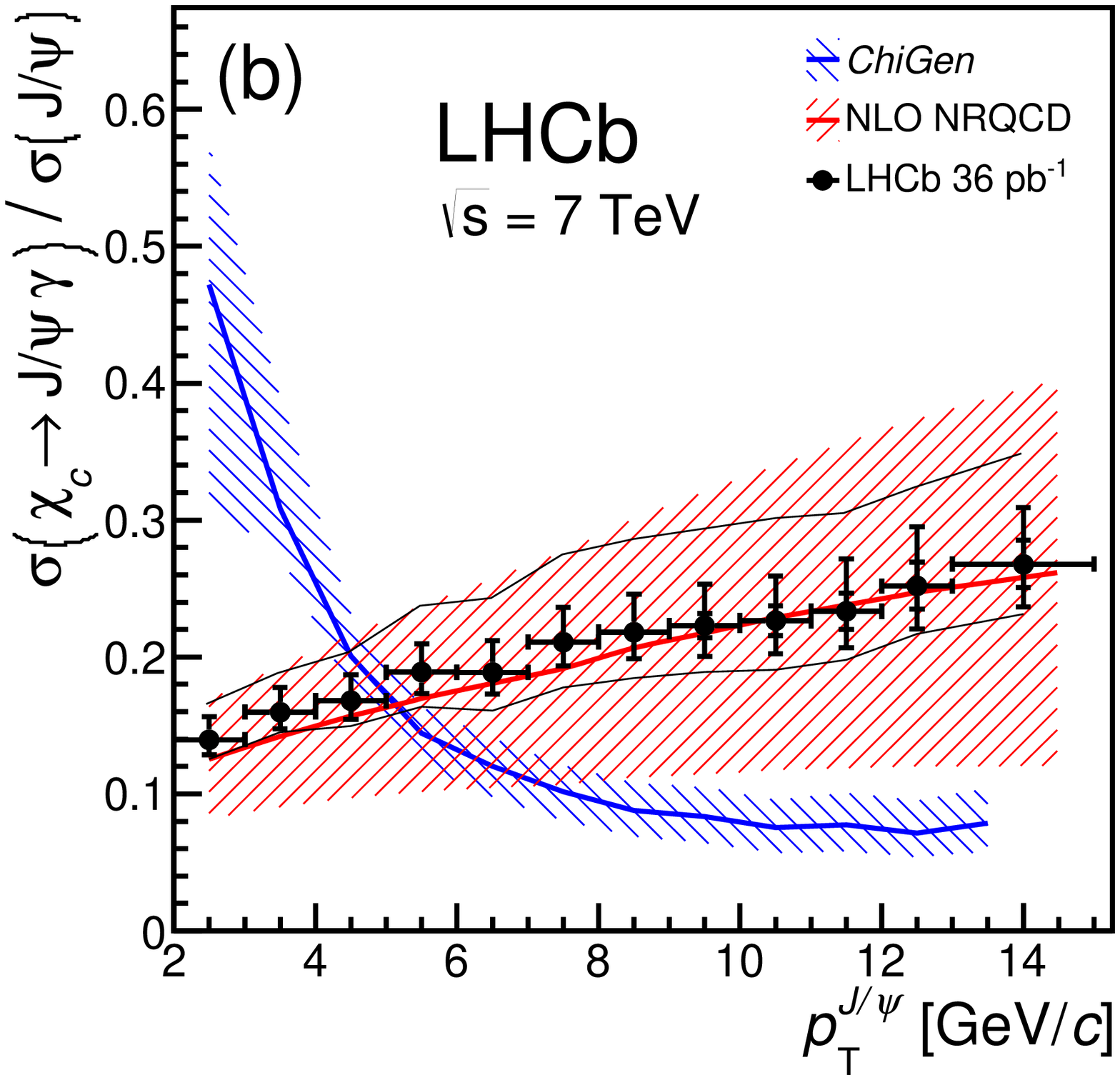}}}
  \end{center}
  \caption{\small{Ratio \srChicToJpsi\ in bins of 
      \pTJpsi\ in the range \pTJpsiRange. 
      The \LHCb\ results,
      in the rapidity range \yRange\ and 
      assuming the production of unpolarised \Jpsi\ and \Chic\ mesons,
      are shown with solid black circles and the internal
      error bars correspond to the statistical error; 
      the external error bars include the contribution 
      from the systematic uncertainties (apart from the polarisation). 
      The lines surrounding the data points show the
      maximum effect of the unknown 
      \Jpsi\ and \Chic\ polarisations on the result. 
      The upper and lower limits correspond to the spin states as described in
      the text. 
      The CDF data points, at
      \ensuremath{\sqrt{s}\myop{=}\myvalue{1.8}{\TeV}} 
      in \ppbar\ collisions and 
      in the \Jpsi\ pseudo-rapidity range \ensuremath{|\eta^{\Jpsi}|<1.0},
      are shown in (a) with open blue circles~\cite{Abe:1997yz}.
      The two hatched bands in (b) correspond 
      to the \chigen\ Monte Carlo generator
      prediction~\cite{Harland:ChiGen} and NLO NRQCD~\cite{Ma:2010vd}.}}
  \label{fig:Results}
\end{figure}

The cross-section ratio, \srChicToJpsi, measured in bins of \pTJpsi\ is
given in \Table~\ref{tab:ResultsEqOne} and shown in \Figure~\ref{fig:Results}.
The measurements are consistent with, 
but suggest a different trend to previous results from CDF
using \ppbar\ collisions at 
\ensuremath{\sqrt{s}=1.8}~TeV~\cite{Abe:1997yz} 
as shown in \Figure~\ref{fig:Results}(a), 
and from HERA-$B$ in \ensuremath{p\mathrm{A}} collisions 
at \ensuremath{\sqrt{s}=41.6}~GeV, 
with \pTJpsi\ below roughly \ensuremath{5}\GeVc, 
which gave
\ensuremath{\srChicToJpsi = 0.188\pm 0.013 ^{+0.024}_{-0.022}}~\cite{Abt:2008ed}.
\begin{table}[tbp]
\renewcommand{\arraystretch}{2}
\caption{
  \label{tab:ResultsEqOne}
  \small{Ratio \srChicToJpsi\ in bins of \pTJpsi\ 
    in the range 
    \pTJpsiRange\ and in the
    rapidity range \yRange.
    The first error is statistical and
    the second is systematic (apart from the polarisation).
    Also given is the maximum effect of the unknown polarisations on the results 
    as described in \Section~\ref{sec:Polarisation}.
}}
\smallskip
\begin{center} 
\begin{tabular}{|c|c|c|} \hline
\pTJpsi (\GeVc)  & \srChicToJpsi\  & Polarisation effects\\ \hline
$2-3$               & $  {0.140}^{+0.005\;+0.015}_{-0.005\;-0.011} $ & $ {}^{+0.025}_{-0.014} $     \\ \hline
$3-4$               & $  {0.160}^{+0.003\;+0.017}_{-0.004\;-0.012} $ & $ {}^{+0.028}_{-0.015} $     \\ \hline
$4-5$               & $  {0.168}^{+0.003\;+0.019}_{-0.003\;-0.012} $ & $ {}^{+0.035}_{-0.018} $     \\ \hline
$5-6$               & $  {0.189}^{+0.004\;+0.021}_{-0.004\;-0.015} $ & $ {}^{+0.048}_{-0.025} $     \\ \hline
$6-7$               & $  {0.189}^{+0.005\;+0.022}_{-0.004\;-0.016} $ & $ {}^{+0.054}_{-0.028} $     \\ \hline
$7-8$               & $  {0.211}^{+0.005\;+0.024}_{-0.005\;-0.017} $ & $ {}^{+0.064}_{-0.033} $     \\ \hline
$8-9$               & $  {0.218}^{+0.007\;+0.026}_{-0.007\;-0.019} $ & $ {}^{+0.068}_{-0.034} $     \\ \hline
$9-10$              & $  {0.223}^{+0.009\;+0.030}_{-0.009\;-0.019} $ & $ {}^{+0.070}_{-0.034} $     \\ \hline
$10-11$             & $  {0.226}^{+0.011\;+0.030}_{-0.011\;-0.022} $ & $ {}^{+0.073}_{-0.036} $     \\ \hline
$11-12$             & $  {0.233}^{+0.013\;+0.034}_{-0.013\;-0.026} $ & $ {}^{+0.070}_{-0.036} $     \\ \hline
$12-13$             & $  {0.252}^{+0.018\;+0.037}_{-0.017\;-0.029} $ & $ {}^{+0.071}_{-0.035} $     \\ \hline
$13-15$             & $  {0.268}^{+0.018\;+0.038}_{-0.017\;-0.025} $ & $ {}^{+0.080}_{-0.037} $     \\ \hline
\end{tabular}
\end{center}
\end{table}

Theory predictions,
calculated in the \lhcb\ rapidity range \yRange, 
from the \chigen\ Monte Carlo generator~\cite{Harland:ChiGen}
and from the NLO NRQCD calculations~\cite{Ma:2010vd}
are shown as hatched bands in \Figure~\ref{fig:Results}(b). 
The \chigen\ Monte Carlo event generator
is an implementation of the leading-order 
colour-singlet model described in \Reference~\cite{Glover:1987az}.
However, since the colour-singlet model implemented in \chigen\
does not reliably predict the prompt \Jpsi\ cross-section,
the \srChicToJpsi\ prediction uses the \Jpsi\ cross-section 
measurement from 
\Reference~\cite{Aaij:2011jh} as the denominator
in the cross-section ratio.  

Figure~\ref{fig:Results} also shows
the maximum effect of the unknown \Jpsi\ and \Chic\
polarisations on the result, shown as lines surrounding the data points. 
In the first \pTJpsi\ bin, 
the upper limit corresponds to a
spin state combination  
\ensuremath{(|m_{\ChicOne}|,
|m_{\ChicTwo}|,
\lambda_{\Jpsi})}
equal to 
\ensuremath{(1,2,-1)} 
and the lower limit to 
\ensuremath{(0,1,1)}.
For all subsequent bins,
the upper and lower limits correspond to the spin state combinations
\ensuremath{(0,2,-1)} 
and
\ensuremath{(1,0,1)} respectively.

In summary, the ratio of the \srChicToJpsi\ prompt production
cross-sections is measured using 36\invpb of data
collected by \LHCb\ during 2010 at a centre-of-mass energy \SqrtS. 
The results provide a 
significant statistical improvement 
compared to previous measurements~\cite{Abt:2008ed,Abe:1997yz}. 
The results are in agreement with the NLO NRQCD model~\cite{Ma:2010vd}
over the full range of \pTJpsi.
However, there is a significant discrepancy compared to the leading-order
colour-singlet model described by the \chigen\ Monte Carlo 
generator~\cite{Harland:ChiGen}.
At high \pTJpsi, NLO corrections fall less slowly with \pTJpsi\
and become important, 
it is therefore not unexpected that the model lies below the data.
At low \pTJpsi, the data appear to put a severe strain on
the colour-singlet model.

\section*{Acknowledgments}
We would like to thank
L.~A.~Harland-Lang, W.~J.~Stirling and K.-T.~Chao for supplying the theory
predictions for comparison to our data and for many helpful discussions.

We express our gratitude to our colleagues in the CERN accelerator
departments for the excellent performance of the LHC. We thank the
technical and administrative staff at CERN and at the LHCb institutes,
and acknowledge support from the National Agencies: CAPES, CNPq,
FAPERJ and FINEP (Brazil); CERN; NSFC (China); CNRS/IN2P3 (France);
BMBF, DFG, HGF and MPG (Germany); SFI (Ireland); INFN (Italy); FOM and
NWO (The Netherlands); SCSR (Poland); ANCS (Romania); MinES of Russia and
Rosatom (Russia); MICINN, XuntaGal and GENCAT (Spain); SNSF and SER
(Switzerland); NAS Ukraine (Ukraine); STFC (United Kingdom); NSF
(USA). We also acknowledge the support received from the ERC under FP7
and the Region Auvergne.

\bibliographystyle{LHCb}
\bibliography{main}

\end{document}